\documentclass[12pt]{article}
\pdfoutput=1
\usepackage{amssymb,amsmath}
\usepackage{color,graphicx}
\usepackage{cite}
\newcommand\fverb{\setbox\pippobox=\hbox\bgroup\verb}
\newcommand\fverbdo{\egroup\medskip\noindent%
                              \fbox{\unhbox\pippobox}\ }
\newcommand\fverbit{\egroup\item[\fbox{\unhbox\pippobox}]}
\newbox\pippobox

\newcommand{\be} {\begin{equation}}
\newcommand{\ee} {\end{equation}}
\newcommand{\beq} {\begin{equation}}
\newcommand{\eeq} {\end{equation}}
\newcommand{\bea} {\begin{eqnarray}}
\newcommand{\eea} {\end{eqnarray}}
\newcommand{\bear}{\begin{eqnarray}}
\newcommand{\eear}{\end{eqnarray}}

\newcommand{\rc}{\nonumber\\}

\def\ie{{\em i.e.}}

\def\om{\omega}

\begin{document}
 
\begin{flushright}
HIP-2016-17/TH
\end{flushright}

\begin{center}

\centerline{\Large {\bf Non-relativistic anyons from holography}}

\vspace{8mm}

\renewcommand\thefootnote{\mbox{$\fnsymbol{footnote}$}}
Niko Jokela,${}^{1,2}$\footnote{niko.jokela@helsinki.fi}
Jarkko J\"arvel\"a,${}^{1,2}$\footnote{jarkko.jarvela@helsinki.fi}
and Alfonso V. Ramallo${}^{3,4}$\footnote{alfonso@fpaxp1.usc.es}

\vspace{4mm}
${}^1${\small \sl Department of Physics} and ${}^2${\small \sl Helsinki Institute of Physics} \\
{\small \sl P.O.Box 64} \\
{\small \sl FIN-00014 University of Helsinki, Finland} 

\vskip 0.2cm
${}^3${\small \sl Departamento de  F\'\i sica de Part\'\i  culas} \\
{\small \sl Universidade de Santiago de Compostela} \\
{\small \sl and} \\
${}^4${\small \sl Instituto Galego de F\'\i sica de Altas Enerx\'\i as (IGFAE)} \\
{\small \sl E-15782 Santiago de Compostela, Spain}

\end{center}

\vspace{8mm}
\numberwithin{equation}{section}
\setcounter{footnote}{0}
\renewcommand\thefootnote{\mbox{\arabic{footnote}}}

\begin{abstract}
\noindent
We study generic types of holographic matter residing in Lifshitz invariant defect field theory as modeled by adding probe D-branes in the bulk black hole spacetime characterized by dynamical exponent $z$ and with hyperscaling violation exponent $\theta$. Our main focus will be on the collective excitations of the dense matter in the presence of an external magnetic field. Constraining the defect field theory to 2+1 dimensions, we will also allow the gauge fields become dynamical and study the properties of a strongly coupled anyonic fluid. We will deduce the universal properties of holographic matter and show that the Einstein relation always holds.
\end{abstract}

\newpage
\tableofcontents

\renewcommand{\theequation}{{\rm\thesection.\arabic{equation}}}


\section{Introduction}

Gauge/gravity duality has achieved its stature in theorists' arsenal to attack problems notoriously difficult to fight with perturbative tools.
This applicability stems from the fact that the duality relates a theory at strong coupling to another theory at weak coupling, and vice versa. This is a particularly useful property when dealing with situations that one would normally describe using gauge field theory techniques, but when such systems are subject to conditions where strong interactions are expected and thus behave drastically differently. The prototypical example is the theory of strong interactions, QCD, at finite baryon chemical potentials. Here the implementation of the gauge/gravity duality, holography, has been successfully utilized both at high \cite{AdS_CFT_reviews} and at low temperatures \cite{Hoyos:2016zke}, natural environments for dense quark matter in heavy ion collisions and at the cores of neutron stars, respectively. 

In its best understood scenario, the gauge/gravity correspondence relates string theory living in an Anti-de Sitter (AdS) spacetime (times a compact manifold) to a conformal field theory (CFT) in one less non-compact spatial dimension. Natural extensions consist of those bulk spacetimes which are still asymptotically AdS and act as dual geometries to relativistic matter. However, in many cases the  configurations one deals with in the laboratories are not relativistic. The bulk geometries then ought to be warped products of Lifshitz  spaces with compact manifolds. However, only a few examples of top-down constructions have been found to possess Lifshitz scaling. Moreover, it seems very subtle to nail down the precise holographic dictionary \cite{Taylor:2015glc}.

While there is no obvious obstruction to deriving generic metrics possessing Lifshitz scaling from string theory, the progress has been excruciatingly slow due to highly technical reasons. For this reason, most of the holographic studies related to Lifshitz geometries have been bottom-up, meaning that some broader form of the gauge/gravity correspondence is assumed while the string theory embedding of the background is lacking. In this paper, we will also follow this approach and start with a background metric possessing Lifshitz scaling with dynamical exponent $z$. We will also allow hyperscaling violation, introduced via an additional parameter $\theta$ in the background metric. The matter in our model is introduced by adding flavor D-branes with appropriate bulk gauge fields turned on in the worldvolume of the brane, in particular, in such a manner that the matter has finite charge density. We note that this has been under systematic study also in the past \cite{HoyosBadajoz:2010kd,Dey:2013vja,Edalati:2013tma}, though essentially only at zero temperature. In this paper, we will also consider thermal effects on the collective excitations of the system, putting special focus on exploring how the system enters in the hydrodynamic regime. We will also discuss charge diffusion and establish the Einstein relation for all parameter values. 

An important new development that we will report is the analysis of the matter in the background of an external magnetic field. The standard prescription of introducing an external magnetic field in holography is via introducing new non-vanishing components for the gauge field living on the brane, $F_{xy}\propto B$. In generic dimensions, one needs to keep $B$ fixed, corresponding to Dirichlet boundary conditions. However, when the bulk spacetime is four-dimensional, an alternative scheme for quantizing the gauge field opens up \cite{Witten:2003ya,Yee:2004ju}. In particular, one can implement combined Dirichlet/Neumann (or Robin) boundary conditions for the gauge field \cite{Jokela:2013hta,Ihl:2016sop}, leading to dynamical gauge fields. In such a scenario, the magnetic field is not kept fixed, but one allows for it to adjust its own expectation value. This leads us to the study of matter which is not only charged electrically, but also carries magnetic charges. These are anyons, particles of fractional statistics, which   are the subject of the latter part of our work. 

There has been a tremendous amount of work devoted to the study of anyons, since their inception in the late seventies \cite{Leinaas:1977fm}. Yet, they are very mysterious and the field is still in its infancy. The main reason for the difficulties arise from the property that multi-anyon states cannot be expressed as a simple product of single particle states. The anyons are linked together via braiding, which might be suggestive of strong interactions. This is precisely where the holography applies and may help in rearranging thoughts in seeking answers to puzzles raised by anyonic fluids. The anyonic fluids have been studied in several holographic works \cite{Jokela:2013hta,Jokela:2015aha,Brattan:2013wya,Brattan:2014moa,Itsios:2015kja,Itsios:2016ffv,Jokela:2014wsa}. The most recent work \cite{Ihl:2016sop} was able to obtain the explicit equation of state for anyons (holographically modeled using a dyonic black brane), an achievement that has been extremely challenging to reach with perturbative methods. Clearly, one should try to implement the prescription given in \cite{Ihl:2016sop} to other setups as well, in particular to those that are presented in this paper. This is, however, beyond the scope of current work.

The collective excitations of our system are dual to the quasinormal modes of the D-brane probes, which are obtained from the fluctuations of the Dirac-Born-Infeld action. At sufficiently high temperature, the system is in a hydrodynamical diffusive regime, characterized by a diffusion constant. We will find a closed expression for this  constant and we will study its dependence on the scaling exponents $z$ and $\theta$, as well as on the magnetic field $B$. At low temperature, the dominant excitation is the so-called holographic zero sound \cite{Karch:2008fa,Karch:2009zz}.  We will determine analytically the dispersion relation of the zero sound in the Lifshitz geometry for non-zero magnetic field.  We will show that the zero sound mode is gapped when $B\not= 0$, generalizing similar previous results in other geometries \cite{Jokela:2012vn,Brattan:2012nb}. We will generalize this analysis to include alternative quantization conditions in the case of $(2+1)$-dimensional field  theories on the boundary. We will find that the effect of the new boundary conditions on the zero sound is similar to the one of a  magnetic field. In particular, we will show that one can adjust  them to make the zero sound gapless, as was found in \cite{Jokela:2015aha,Itsios:2015kja,Itsios:2016ffv} for relativistic backgrounds. 
We will also study the diffusion constant and the conductivities of the anyonic fluid.

The organization of the rest of this paper is the following. 
We will begin by introducing the background geometry in section \ref{setup}. We embed a probe D-brane in a generic black hole metric possessing both Lifshitz scaling with dynamical exponent $z$ and hyperscaling violating exponent $\theta$. We review the basic thermodynamic properties and then focus on deriving the fluctuation equations extracted from perturbing the flavor brane embedding function and the gauge fields. Section \ref{sec:collective} solves the fluctuation equations in various limits of the background charge density, magnetic field, and temperature. We also compare the analytic results that we will obtain to those from numerics. In section \ref{sec:alternative}, we switch gears by constraining to 2+1 dimensions to allow the gauge fields become dynamical via alternative quantization. We will discuss collective excitations of the resulting anyonic fluid. One important result that we can establish is the Einstein relation in the most generic case. This result we will take literally in section \ref{sec:comments} and predict the conductivity of the matter at finite (and large) magnetic field strength, a regime where some of the other approximation schemes fall short. Section \ref{conclusions} contains a brief summary of our results together with remarks on a few open problems left in future works.
The paper is complemented with several appendices which contain technical steps filling in the gaps in the calculations of the bulk text.


\section{Set-up}\label{setup}

We begin by introducing the bulk geometry which exhibits Lifshitz scaling and hyperscaling violation. We then embed flavor probe D-brane in this background, thus introducing massless quenched fundamental degrees of freedom localized on a lower-dimensional defect field theory with flavor symmetry $U(N_f)$. We analyze the thermodynamics of such holographic matter at non-zero baryonic charge density by introducing a chemical potential for the diagonal $U(1)\subset U(N_f)$. 

\subsection{Background}

Let us consider the following $(p+2)$-dimensional metric,
\bea
ds^2_{p+2} & = & g_{tt}(r)f_p(r) dt^2+g_{xx}(dx^i)^2+\frac{g_{rr}}{f_p(r)}dr^2 \nonumber \\
           & = & r^{-\frac{2\theta}{p}}\left[-f_p(r)r^{2z}dt^2+r^2 (dx^i)^2 +\frac{dr^2}{f_p(r) r^2}\right] \ , \ i=1,\ldots,p \ , \label{eq:metric}
\eea
where the blackening factor reads
\be\label{eq:blackening}
 f_p = 1-\left(\frac{r_h}{r}\right)^{p\,\xi + z} \ ,
\ee
and where the metric components we record separately for ease of reference
\be
 g_{tt}(r) = -r^{2\xi+2(z-1)} \qquad , \qquad g_{xx}(r)=r^{2\xi} \qquad , \qquad g_{rr}(r)=r^{2\xi-4} \ .
\ee
In (\ref{eq:blackening}) we have defined a parameter $\xi$, which is related to the hyperscaling violating parameter $\theta$ as follows:
\be
 \xi = 1-\frac{\theta}{p} \ ,
\ee
and $z$ is the dynamical exponent.
We note that the radial coordinate $r$ is defined as is standard, \ie, $r=\infty$ corresponds to the boundary, where the field theory lives and $r_h$ is the horizon radius of the black hole. By demanding the absence of conical singularity in the bulk, the horizon radius can be related to the field theory temperature according to
\be
 r_h = \left(\frac{4\pi T}{p+z-\theta}\right)^{\frac{1}{z}} \ .
\ee
Here we are assuming that 
\be
 z\geq 1 \quad , \quad \theta\leq 0 \ .
\ee
Realizations of $z<1$ seem pathological as they lead to violations of the null energy condition \cite{Hoyos:2010at}, whereas the latter requirement comes from thermodynamic stability (see below).

We now wish to embed $N_f$ probe D-branes in this background, so that the branes are extended in $q\leq p$ spatial dimensions of the Lifshitz spacetime (\ref{eq:metric}). In the generic case then the flavor fields reside on a ($q+1$)-dimensional defect. We will consider the following ansatz for the gauge field on the probes:
\be\label{eq:F}
 F = A'_t dr\wedge dt + B dx^1\wedge dx^2 \ ,
\ee
where the prime denotes a derivative with respect to $r$. The Dirac-Born-Infeld (DBI) action for massless probes thus reads
\be\label{eq:DBIaction}
 S = -N_f T_{D}V\int dt dr d^{q}x\sqrt{-\det\left(g+F\right)} =  -{\cal N}\int dr \sqrt H\sqrt{|g_{tt}|g_{rr}-A'^2_t} \ ,
\ee
where ${\cal N} = N_f T_{D}V_{q+1}V$, $T_D$ is the tension, $V$ is the volume of the internal space which the D-branes may be wrapping, and the function $H$ is
\be
 H = g_{xx}^2+g_{xx}^{q-2}B^2 = r^{2q\xi}+r^{2(q-2)\xi}B^2 \ .
\ee
Notice that we have not included a dilaton which is generically non-trivial in top-down string theory constructions dual to non-conformal field theories. 

The equation of motion for $A_t$, which follows from (\ref{eq:DBIaction}), can be integrated once to find
\be\label{eq:At_prime}
 A'_t = \frac{d\sqrt{g_{rr}|g_{tt}|}}{\sqrt{H+d^2}} \ ,
\ee
where $d$ is an integration constant, proportional to the physical charge density of the field theory: $\langle J^t\rangle\equiv {\cal N}d$. From now on, we will consider $d$ to be positive.

\subsection{Thermodynamics}

Let us now proceed with discussing some properties of the probe brane system. We are, in particular, interested in thermodynamic relations and how the parameters $z$, $\xi$, and $q$ affect them.

\paragraph{Zero temperature}\mbox{}\newline

Let us first consider the system at $T=0$ with vanishing magnetic field $B=0$. From (\ref{eq:At_prime}) we have that:
\beq
A_t'\,=\,d\,{r^{2\xi+z-3}\over \sqrt{r^{2q\xi}+d^2}}\,\,\,.
\eeq
Therefore, the (zero-temperature) chemical potential is:
\beq\label{eq:chem0a}
\mu_0=A_t(\infty)=\int_0^{\infty}\,dr\,A_t'\,=\,d\,\int_0^{\infty}\,{r^{2\xi+z-3}\over \sqrt{r^{2q\xi}+d^2}}dr \equiv d\,I_{2\xi+z-3\,,\,2q\xi}(r=0) \ .
\eeq
The integrals of the kind (\ref{eq:chem0a}) appear frequently in this paper, for sake of which we have defined two classes of integrals and collected their useful properties in Appendix \ref{appendix:calculations}.
The explicit form of the chemical potential follows
\beq
\mu_0\,=\,\gamma\,d^{{2\xi+z-2\over \xi q}}\qquad , \qquad \gamma\,=\,{1\over 2\xi q}\,\, B\Big({2\xi+z-2\over 2\xi q}\,,\,{\xi (q-2)+2-z\over 2\xi q}\Big) \ .
\eeq
The on-shell action is:
\beq
S_{on-shell}\,=\,-{\cal N}\,\int_0^{\infty}\,dr\,
{\sqrt{g_{rr}|g_{tt}|}\over \sqrt{H+d^2}}\,\,H\,=\,
-{\cal N}\,\int_0^{\infty}\,{r^{2\xi q+2\xi+z-3}\over 
\sqrt{r^{2 q\xi}+d^2}}\,dr\,\,.
\eeq
This is a divergent integral. We regulate it by subtracting an on-shell action for probe branes at zero density:
\beq
S_{on-shell}^{reg}\,=\,-{\cal N}\,\int_0^{\infty}\,dr\,
r^{\xi q+2\xi+z-3}\Bigg[ {r^{\xi q}\over 
\sqrt{r^{2 q\xi}+d^2}}\,-\,1\Bigg]\,\,.
\eeq
To evaluate this integral we use the general result:
\beq
\int_0^{\infty}\,r^{{\lambda_2\over 2}}\,\Big[{r^{{\lambda_1\over 2}}\over 
\sqrt{r^{\lambda_1}+d^2}}\,-1\Big]\,dr\,=\,
{1\over \lambda_1}\,B\Big(-{\lambda_2+2\over 2\lambda_1}\,,\,
{1\over 2}+{\lambda_2+2\over 2\lambda_1}\Big)\,
d^{{\lambda_2+2\over \lambda_1}}\,\,,
\eeq
which is valid for $\lambda_2<2(\lambda_1-1)$. We get:
\beq
S_{on-shell}^{reg}\,=\,-{{\cal N}\over 2 q\xi}\,
B\Big(-{q\xi+2\xi+z-2\over 2 q\xi}\,,\,
{2q\xi+2\xi+z-2\over 2 q\xi}\Big)\,d^{1+{2\xi+z-2\over q\xi}}\,\,.
\eeq
The (zero temperature) grand potential $\Omega_0=\Omega_0(\mu_0)=-S_{on-shell}^{reg}$ reads,
\beq\label{eq:OmatT0}
\Omega_0\,=\,-{2\xi+z-2\over q\xi+2\xi+z-2}\,{\cal N}\,\gamma\,d^{1+{2\xi+z-2\over q\xi}} =-{2\xi+z-2\over q\xi+2\xi+z-2}\,{\cal N}\,
\gamma^{-{\xi q\over 2\xi+z-2}}\,\mu_0^{1+{\xi q\over 2\xi+z-2}} \ .
\eeq
It is now straightforward to obtain the density $\rho=\langle J^t\rangle$ as:
\beq
\rho\,=\,-{\partial\Omega_0\over \partial\mu_0}\,=\,{\cal N}\,d\,\,,
\label{density}
\eeq
\ie, $d$ is proportional to $\rho$, as promised. The energy density can be obtained by Legendre transformation $\epsilon\,=\,\Omega_0+\mu_0\,\rho$:
\beq
\epsilon\,=\,{q\xi\over q\xi+2\xi+z-2}\,{\cal N}\,\gamma\,
d^{1+{2\xi+z-2\over q\xi}}\,\,.
\eeq
For the pressure we find:
\beq
P = -\Omega_0 ={2\xi+z-2\over q\xi+2\xi+z-2}\,{\cal N}\,\gamma\,d^{1+{2\xi+z-2\over q\xi}} \,=\,
{2\xi+z-2\over q\xi}\,\epsilon\,\,.
\eeq
Therefore, the speed of first sound is:
\beq
u_s^2\,=\,{\partial P\over \partial\epsilon}\,=\,
{2\xi+z-2\over q\xi}\,\,. \label{eq:speed-1st-sound}
\eeq
This result agrees with the one found in \cite{Edalati:2013tma}.

\paragraph{Non-zero temperature}\mbox{}\newline

To extract more useful information, we commit to heat up the system.
Let us begin by analyzing the chemical potential at $T\not=0$:
\beq\label{eq:chem}
\mu= d\,\int_{r_h}^{\infty}\,{r^{2\xi+z-3}\over \sqrt{r^{2q\xi}+d^2}}dr = \mu_0-{r_h^{2\xi+z-2}\over 2\xi+z-2}\,
F\Big({1\over 2}, {2\xi+z-2\over 2 q\xi}; 1+{2\xi+z-2\over 2 q\xi};-{r_h^{2 q\xi}\over d^2}\Big) \ , 
\eeq
where $\mu_0$ is the chemical potential at zero temperature (\ref{eq:chem0a}). 
The grand potential at $T\not=0$ is:
\beq
\Omega\,=\,{\cal N}\,\int_{r_h}^{\infty}\,dr\,
r^{\xi q+2\xi+z-3}\Bigg[ {r^{\xi q}\over 
\sqrt{r^{2 q\xi}+d^2}}\,-\,1\Bigg]\,\,.
\eeq
We evaluate this integral using the formula:
\bear
&&\int_{r_h}^{\infty}\,r^{{\lambda_2\over 2}}\,\Big[{r^{{\lambda_1\over 2}}\over 
\sqrt{r^{\lambda_1}+d^2}}\,-1\Big]\,dr\,=\,
{1\over \lambda_1}\,B\Big(-{\lambda_2+2\over 2\lambda_1}\,,\,
{1\over 2}+{\lambda_2+2\over 2\lambda_1}\Big)\,
d^{{\lambda_2+2\over \lambda_1}}\,+\,{2\over 2+\lambda_2}\,r_h^{{\lambda_2+2\over 2}} \rc
&&
\qquad\qquad\qquad
-{2\over \lambda_1+\lambda_2+2}\,{r_h^{1+{\lambda_1+\lambda_2\over 2}}}\,
F\Big({1\over 2}, {2+\lambda_1+\lambda_2\over 2\lambda_1}; 
{2+3\lambda_1+\lambda_2\over 2\lambda_1};-{r_h^{\lambda_1}\over d^2}\Big)\,\,.
\eear
We find
\beq
\Delta\Omega\,=\,\Omega_0\,-\,{{\cal N}\over 2 q\xi+2\xi+z-2}\,{r_h^{2 q\xi+2\xi+z-2}\over d}\,
F\Big({1\over 2}, 1+{2\xi+z-2\over 2 q \xi}; 
2+{2\xi+z-2\over 2 q \xi};-{r_h^{2 q\xi}\over d^2}\Big)\,\,,
\eeq
where $\Omega_0$ is the grand potential at zero temperature (\ref{eq:OmatT0}) and $\Delta\Omega$ is the density-dependent part of $\Omega$, defined as:
\beq
\Delta\Omega\,=\,\Omega-{r_h^{q\xi +2\xi+z-2}\over q\xi +2\xi+z-2}\,\,.
\eeq
Notice that the natural variable of $\Omega$ is $\mu$, as $d$ depends on it through (\ref{eq:chem}). At low temperature we can explicitly invert (\ref{eq:chem}). Indeed, let us consider the low temperature case in which $r_h$ is small. The chemical potential can then be expanded as:
\beq\label{eq:muinvert}
\mu = \mu_0\,-\,{r_h^{2\xi+z-2}\over 2\xi+z-2}\,+\,{1\over 2}\,
{1\over 2q\xi +2\xi+z-2}\,{r_h^{2q\xi +2\xi+z-2}\over d^2}\,+\,\ldots \ .
\eeq
The expansion of $\Delta \Omega$ is:
\beq
\Delta\Omega=-{2\xi+z-2\over q\xi+2\xi+z-2}\,{\cal N}\,\gamma\,d^{1+{2\xi+z-2\over q\xi}}\,-\,
{{\cal N}\over 2q\xi +2\xi+z-2}\,{r_h^{2q\xi +2\xi+z-2}\over d}\,+\,
\ldots\,\,.
\eeq
Plugging in the expression for $d=d(\mu)$ from (\ref{eq:muinvert}), we can write at leading order in temperature:
\beq
\Delta\Omega\,=\,
-{2\xi+z-2\over q\xi+2\xi+z-2}\,{\cal N}\,\gamma^{-{q\xi\over 2\xi+z-2}}\,
\Big[\mu\,+\,{r_h^{2\xi+z-2}\over 2\xi+z-2}\Big]^{{q\xi\over 2\xi+z-2}}\,+\,\ldots \ .
\eeq
Let us then compute the entropy
\beq
s=-{\partial \Omega\over \partial T}\Big|_{\mu}\,=\,-
{\partial \Omega\over \partial r_h}\Big|_{\mu}\,
{\partial r_h\over \partial T}\,\,.
\eeq
After some calculation we get (at low temperature):
\beq
s\,\approx\,{{\cal N}\,q(p-\theta)\,\over  z(p(q+z)-(2+q)\theta) }\,\frac{d}{\mu}\,\Big[{4\pi\over p+z-\theta}\Big]^{1-{2\theta\over p z}}\,\,T^{-{2\theta\over pz}}\,\,.
\eeq
Notice that the $T$ behavior coincides with the one found in \cite{Dey:2013vja} but the coefficient is different. The specific heat at low temperature thus scales as:
\beq
c_v\,=\,T\,{\partial s\over \partial T}\Big|_{d}\,\sim\, T^{-{2\theta\over pz}} \ .
\eeq
The stability of the system (\ie, $c_v\geq 0$) then requires $\theta/z\leq 0$.

\subsection{Fluctuations}
\label{fluctuation_equations}

Having understood the thermodynamics of the underlying holographic fluid and the scaling upon varying $\xi$ and $z$, we now wish to lay out a framework to exploring the response of the fluid under small perturbations. The relevant physics we are after are due to vector (gauge) fluctuations; the scalar deformations turn out to decouple as we focus on massless flavor degrees of freedom. We will thus consider fluctuations of the form:
\bea
 A = A^{(0)}+a(r,x^\mu) \ ,
\eea
where $A^{(0)} = A^{(0)}_\nu dx^\nu = A_t dt+ B x^1 dx^2$ and $a(r,x^\mu)=a_\nu(r,x^\mu)dx^\nu$. The total gauge field strength is:
\be
 F = F^{(0)}+f \ ,
\ee
where $F^{(0)}= dA^{(0)}$ is the two-form written in (\ref{eq:F}) and $f=da$. We note that we consider the fluctuations to depend only on $r,t,x^1$ as we can always choose the momentum vector to align along one of the spatial directions. 

The powerful method to fluctuating the DBI action is the approach introduced in \cite{Jokela:2015aha}. In fact, we can just quote the corresponding results in \cite{Jokela:2015aha} by first stating the relevant elements of the open string metric:
\bear
{\cal G}^{tt} & = & -{1\over f_p}\,{g_{rr}\over g_{rr}\,|g_{tt}|\,-\,A_t'^{\,2}}\,=\,
-{H+d^2\over f_p\,|g_{tt}|\,H}\rc
{\cal G}^{r r} & = & f_p\,{|g_{tt}|\over g_{rr}\,|g_{tt}|\,-\,A_t'^{\,2}}\,=\,
{H+d^2\over g_{rr}\,H}\,f_p \rc
{\cal G}^{x^1\,x^1} & = & {\cal G}^{x^2\,x^2}\,=\,{g_{xx}\over g_{xx}^2+B^2} \ ,
\eear
while those of the antisymmetric matrix ${\cal J}$ are:
\bear
{\cal J}^{tr} & = & -{\cal J}^{r t}\,=\,-{A_t'\over g_{rr}\,|g_{tt}|\,-\,A_t'^{\,2}}\,=\,
-{d\over \sqrt{|g_{tt}|\,g_{rr}}}\,
{\sqrt{H+d^2}\over H}
\rc
{\cal J}^{x^1\,x^2} & = & -{\cal J}^{x^2\,x^1}\,=\,-{B\over g_{xx}^2+B^2}\,\,.
\eear
The Lagrangian for the fluctuations is:
\beq\label{eq:fluctuationLagrangian}
{\cal L}\,\sim\,{\sqrt{g_{rr}\,|g_{tt}|}\over \sqrt{H+d^2}}\,\,H\,\,
\Big(\,{\cal G}^{ac}\,{\cal G}^{bd}\,-\,
{\cal J}^{ac}\,{\cal J}^{bd}\,+\,{1\over 2}\,{\cal J}^{cd}\,{\cal J}^{ab}
\Big)f_{cd}\,f_{ab} \ ,\ \  \ a,b,c,d\in\{t,x,y,r\} \ .
\eeq
The corresponding equation of motion for $a_d$ then follows:
\beq
\partial_{c}\,\Bigg[{\sqrt{g_{rr}\,|g_{tt}|}\over \sqrt{H+d^2}}\,\,H
\Big(\,{\cal G}^{ca}\,{\cal G}^{db}\,-\,
{\cal J}^{ca}\,{\cal J}^{db}\,+\,{1\over 2}\,{\cal J}^{cd}\,{\cal J}^{ab}
\Big)\,f_{ab}\Bigg]\,=\,0\,\,.
\label{eom_general}
\eeq
From the equation of motion for $a_{r}$ (with $a_{r}=0$) we get the  transversality condition:
\beq
\partial_t\,a_t'\,-\,u^2(r)\,\partial_x\,a_x'\,=\,0\,\,,
\label{transversality_xt}
\eeq
where $u(r)$ is the function:
\beq
u^2(r)\,=\,-{{\cal G}^{xx}\over {\cal G}^{tt}}=\,{g_{xx}\,|g_{tt}|\,f_p\over g_{xx}^2\,+B^2}\,\,{H\over H+d^2} = {f_p r^{2q\xi+2\,z-2}\over 
r^{2q\xi}\,+\,r^{2(q-2)\xi}\,B^2\,+\,d^2} \ .
\eeq
The next step is to Fourier transform the fields:
\be
 a_\nu(r,t,x) = \int \frac{d\omega dk}{(2\pi)^2}a_\nu(r,\omega,k)e^{-i\omega t+ikx} \ .
\ee
We now define the electric field $E$ as the gauge-invariant combination:
\beq
E\,=\,k\,a_t\,+\,\omega\,a_x\,\,.
\label{E_at_ax}
\eeq
Using the transversality condition (\ref{transversality_xt}) in the momentum space, we obtain $a_t'$ and $a_x'$ in terms of $E'$ as follows:
\beq
a_t'\,=\,-{k\,u^2\over \omega^2\,-\,k^2\,u^2}\,E'\,\,,
\qquad\qquad
a_x'\,=\,{\omega\over \omega^2\,-\,k^2\,u^2}\,E'\,\,.
\label{at_ax_E}
\eeq
Using these relations, we obtain the equation of motion for the electric field $E$, as follows:
\bear
&&E''\,+\,\partial_{r}\log\Bigg[
{\sqrt{|g_{tt}|}\over \sqrt {g_{rr}}}\,{g_{xx}\,f_p\over  g_{xx}^2+B^2}\,
{\sqrt{H+d^2}\over \omega^2-k^2\,u^2}\,\Bigg]\,E'\,+\,
{g_{rr}\over |g_{tt}|\,f_p^2}\,(\omega^2-k^2\,u^2)\,E\rc
&&\qquad\qquad\qquad\qquad
=i\,B\,d\,{\sqrt{g_{rr}}\over \sqrt{|g_{tt}|}}\,{g_{xx}^{\,2}+B^2\over g_{xx}\,f_p}\,
{\omega^2-k^2\,u^2\over \sqrt{H+d^2}}\,
\partial_{r}\Bigg({1\over g_{xx}^2+B^2}\Bigg)\,a_y \ .
\eear
Let us write more explicitly this expression by plugging in the value of the function $u$ and the following relation:
\beq
\omega^2\,-\,k^2\,u^2\,=\,
{(\omega^2-r^{2z-2}\,f_p\, k^2)r^{2\xi q}+\omega^2\,B^2\,r^{2\xi (q-2)}+\omega^2\,d^2\over 
r^{2\xi q}+r^{2\xi (q-2)} B^2+d^2}\,\,.
\eeq
The equation for the fluctuation of the electric field $E$ becomes:
\bear\label{eq:Eeom}
&&E''+\,\partial_r\log\Bigg[
{r^{2\xi+z+1}\over r^{4\xi}+B^2}\,
{(r^{2\xi q}+r^{2\xi (q-2)} B^2+d^2)^{{3\over 2}} f_p\over
(\omega^2-r^{2z-2}\,f_p\, k^2)r^{2\xi q}+\omega^2\,B^2\,r^{2\xi (q-2)}+\omega^2\,d^2}
\Bigg]\,E'\rc
&&\qquad\qquad
+{1\over r^{2z+2}f_p^2}\,{
(\omega^2-r^{2z-2}f_p\, k^2)r^{2\xi q}+\omega^2\,B^2\,r^{2\xi (q-2)}+\omega^2\,d^2\over 
r^{2\xi q}+r^{2\xi (q-2)} B^2+d^2}\,E\rc
&&=-4i\xi Bd\,{r^{2\xi-z-2}\over (r^{4\xi}+B^2)f_p}\,
{(\omega^2-r^{2z-2}f_p\, k^2)r^{2\xi q}+\omega^2\,B^2\,r^{2\xi (q-2)}+\omega^2\,d^2\over 
(r^{2\xi q}+r^{2\xi (q-2)} B^2+d^2)^{{3\over 2}}}\,a_y\,\,.
\eear
Moreover, the equation for $a_y$ can be written as:
\bear
&&a_y''\,+\,\partial_{r}\,\log\,\Bigg[{g_{xx}\,\sqrt{|g_{tt}|}\over \sqrt{g_{rr}}}\,f_p\,
{\sqrt{H+d^2}\over g_{xx}^2+B^2}
\Bigg]\,a_y'\,+\,
{g_{rr}\over f_p^2\,|g_{tt}|}\,
(\omega^2-k^2\,u^2)\,a_y \nonumber\\
&&\qquad\qquad\qquad\qquad
\,=\,-iB\,d\,{\sqrt{g_{rr}}\over g_{xx}\,\sqrt{|g_{tt}|}}\, {1\over f_p}\,
{g_{xx}^2+B^2\over \sqrt{H+d^2}}\,
\partial_{r}\Bigg({1\over g_{xx}^2+B^2}\Bigg)\,E
\,\,. 
\eear
Using the expressions for $u$ and the metric elements, this equation becomes:
\bear\label{eq:ayeom}
&&a_y''\,+\,\partial_r\log\Bigg[{r^{2\xi+z+1}\over r^{4\xi}+B^2}\,
(r^{2\xi q}+r^{2\xi (q-2)} B^2+d^2)^{{1\over 2}} f_p\Bigg]\,a_y'\rc
&&\qquad\qquad
+{1\over r^{2z+2}f_p^2}\,{
(\omega^2-r^{2z-2}f_p\, k^2)r^{2\xi q}+\omega^2\,B^2\,r^{2\xi (q-2)}+\omega^2\,d^2\over 
r^{2\xi q}+r^{2\xi (q-2)} B^2+d^2}\,a_y\rc
&&\qquad\qquad
=4i\xi Bd\,{r^{2\xi-z-2}\over (r^{4\xi}+B^2)f_p}\,
{E\over (r^{2\xi q}+r^{2\xi (q-2)} B^2+d^2)^{{1\over 2}}}\,\,.
\eear

The rest of the paper analyzes the solutions to (\ref{eq:Eeom}) and (\ref{eq:ayeom}) in different regimes, at high temperature in section \ref{sec:diffusion} and at low temperature in section \ref{sec:zerosound}. A particularly interesting setting can be obtained in the special case of $q=2$ as one can make use of mixed boundary conditions for the gauge fluctuations. This leads us to the study of anyons and is the topic of section \ref{sec:alternative}.

\subsection{Scalings}
Let us now see how one can eliminate $r_h$ in the equations of motion by rescaling. First of all, we define the new rescaled radial coordinate $\hat r$ as:
\beq
r\,=\,r_h\,\hat r\,\,.
\eeq
This rescaling eliminates $r_h$ from the blackening factor $f_p$. Moreover, it is easy to see that the different factors in (\ref{eq:Eeom}) and (\ref{eq:ayeom}) transform homogeneously if $\omega$, $k$, $d$, and $B$ are rescaled as:
\beq
\omega\,=\,r_h^{z}\,\hat \omega\,\,,
\qquad\qquad
k\,=\,r_h\,\hat k\,\,,
\qquad\qquad
d\,=\,r_h^{\xi q}\,\hat d\,\,,
\qquad\qquad
B\,=\,r_h^{2\xi}\,\hat B \ .\label{eq:scalings1}
\eeq
Notice that $\omega$ and $k$ are rescaled differently for $z\ne 1$, in agreement with the Lifshitz nature of the metric. For this same reason the different components of the gauge field $a_{\mu}$ must transform differently. As the fluctuation equations are linear, we can simply assume that the electric field $E$ does not transform. As $E=\omega a_x+k a_t$, it is clear that $a_t$ and $a_x$ should be rescaled as:
\beq
a_t\,=\,r_h^{-1}\,\hat a_t\,\,,
\qquad\quad
a_x\,=\,r_h^{-z}\,\hat a_x\,\,.
\eeq
Due to symmetry of the indices, $a_y$ should transform as $a_x$. Thus:
\beq
a_y\,=\,r_h^{-z}\,\hat a_y\,\,.
\eeq
It is now straightforward to verify that the two equations of motion scale homogeneously and that working with the hatted variables is equivalent to taking $r_h=1$. Of course, instead of the temperature, one could have chosen to scale out $d$ or $B$, too.


\section{Collective excitations}\label{sec:collective}

In this section, we will analyze the collective excitations of the magnetized brane probes in the Lifshitz background. These collective excitations are dual to the quasinormal modes of the fluctuation equations of section \ref{fluctuation_equations}. We  consider first the system at non-zero temperature and we will look for hydrodynamic diffusive modes. By employing analytical techniques, we obtain the expression of the diffusion constant, which we compare with the result obtained from the numerical integration of the fluctuation equations. We then consider the system at zero temperature and find the dispersion relation of the zero sound mode in the collisionless regime.  Again, we obtain analytic results which we then compare with the  numerical values. The transition between the collisionless and hydrodynamic regime is studied numerically. 

\subsection{Diffusion constant}\label{sec:diffusion}
Let us start by analyzing the equation (\ref{eq:Eeom}) for the fluctuation of the electric field $E$ near the horizon $r=r_h$. 
The blackening factor $f_p(r)$ behaves near  $r=r_h$ as:
\beq
f_p={z+p\,\xi\over r_h}(r-r_h)\,+\,\ldots\,\,.
\eeq
The coefficients of $E$ and $E'$  in (\ref{eq:Eeom}) can be expanded near $r=r_h$ as:
\bear
&&\partial_r\log\Bigg[
{r^{2\xi+z+1}\over r^{4\xi}+B^2}\,
{(r^{2\xi q}+r^{2\xi (q-2)} B^2+d^2)^{{3\over 2}} f_p\over
(\omega^2-r^{2z-2}\,f_p\, k^2)r^{2\xi q}+\omega^2\,B^2\,r^{2\xi (q-2)}+\omega^2\,d^2}
\Bigg] = \frac{1}{r-r_h}+c_1+\ldots\qquad\rc\rc
&&{1\over r^{2z+2}f_p^2}\,{
(\omega^2-r^{2z-2}f_p\, k^2)r^{2\xi q}+\omega^2\,B^2\,r^{2\xi (q-2)}+\omega^2\,d^2\over 
r^{2\xi q}+r^{2\xi (q-2)} B^2+d^2} = \frac{A}{(r-r_h)^2}+\frac{c_2}{r-r_h}\,+\,\ldots \ ,
\eear
where $A$, $c_1$, and $c_2$ are the following constant coefficients:
\bear
 A &=& \frac{\omega ^2\, r_h^{-2 z}}{(\xi  p+z)^2} \rc
c_1 &=& \frac{ r_h^{2 z-3} (\xi  p+z)}{ \left( B^2\, r_h^{-4 \xi }+d^2 r_h^{-2 \xi  q}+1\right)}\,
{k^2\over \omega^2}
 \rc\rc
&&+\,  {\xi\over r_h}\,\Bigg[{(q-2)\,B^2  r_h^{-4 \xi }+q\over B^2\, r_h^{-4 \xi }+d^2\, r_h^{-2 \xi  q}+1}
-{4\over B^2\, r_h^{-4 \xi }+1}\Bigg]\,+\,{z+1+(4-p)\xi\over 2 r_h}
\rc\rc
c_2 &=& -\frac{ r_h^{-3}}{(\xi  p+z) \left( B^2\, r_h^{-4 \xi }+d^2\,r_h^{-2 \xi  q}+1\right)}
\,k^2\,-\,\frac{ r_h^{-2 z-1} (-\xi  p+z+1)}{(\xi  p+z)^2}\,\omega ^2.
\eear
We want to solve (\ref{eq:Eeom}) in the hydrodynamic  diffusive regime in which $k$ and $\omega$ are small and
$\omega={\mathcal O}(k^2)={\mathcal O}(\epsilon^2)$. In this regime, we can neglect the right-hand side of (\ref{eq:Eeom}) and the equations for $E$  and $a_y$ decouple.  Near the horizon we solve (\ref{eq:Eeom}) in a Frobenius series of the type:
\beq\label{Frobenius_expansion}
E = E_{nh}(r-r_h)^{\alpha}\,\Big[1+\beta (r-r_h)\,+\,\ldots\Big] \ ,
\eeq
where $E_{nh}$ is a constant. One can easily show that the exponent $\alpha$  in (\ref{Frobenius_expansion}) is given by:
\beq
\alpha\,=\,-i\,{\omega\over (z+p\,\xi)\,r_h^z}\,\,, \label{eq:alpha}
\eeq
whereas the coefficient $\beta$, at leading order in $\epsilon$,  is given by:
\beq
\beta\,\approx \,-\alpha\,c_1\,=\,i\,{k^2\over \omega}\,{r_h^{z-3}\over  1+B^2r_h^{-4 \xi }+d^2r_h^{-2 \xi  q}} \ .
\eeq
Notice from (\ref{eq:alpha})  that $\alpha\sim {\mathcal O}(\epsilon^2)$ and, therefore, we can neglect the 
$(r-r_h)^{\alpha}$ prefactor in (\ref{Frobenius_expansion}). Thus we write:
\beq
E\approx E_{nh}\Big[1+\beta (r-r_h)\Big]\,\,.
\label{nh_lowfreq_diff_expansion}
\eeq
We now analyze (\ref{eq:Eeom}) by taking the limit at low frequencies first.  In this limit,  we can neglect the terms without derivatives and (\ref{eq:Eeom}) becomes:
\beq
E''\,+\,\partial_{r}\log \left(\frac{r^{\xi(q-2)-z+3} \left(1+B^2 r^{-4\xi}+d^2 r^{-2\xi q }\right)^{3/2}}{1+B^2r^{-4 \xi}}\right)\,E'=0 \ ,
\eeq
which can be readily integrated to give
\beq
E = E^{(0)} + c_E \int\limits_{r}^{\infty}\frac{\left(1+B^2\rho ^{-4 \xi }\right) \rho ^{\xi(2-q) +z-3}}{\left(1+
B^2\rho ^{-4 \xi }+d^2\rho^{-2\xi q}\right)^{3/2}} d\rho\,\,,
\label{E_low_freq_diff}
\eeq
where $E^{(0)}$ and $c_E$ are constants. Notice that $E^{(0)}=E(r\to \infty)$. 
The integral in (\ref{E_low_freq_diff})  does not have a closed analytic form in general but we can easily study its properties in the UV and IR limits. Near the horizon, it has the form
\beq
E = E^{(0)}\,+\,c_E\, \mathcal{I}\, -\,c_E\,\frac{\left(1+B^2r_h^{-4\xi}\right) r_h ^{\xi(2-q)+z-3}}{\left( 1+B^2r_h ^{-4 \xi}+d^2r_h^{-2 \xi q}\right)^{3/2}}\,(r-r_h)\,+\,\ldots\,\,,
\label{E_low_freq_nh_diff}
\eeq
where $\mathcal{I}$ is defined as the integral:
\beq
\mathcal{I} = \int\limits_{r_h}^{\infty}\frac{\left(1+B^2\rho^{-4 \xi }\right) \rho ^{\xi(2-q) +z-3}}{\left( 1+ B^2\rho^{-4\xi}+d^2\rho ^{-2 \xi  q} \right)^{3/2}} d\rho. 
\label{math_I_def}
\eeq
Let us switch to the rescaled variables $\hat{d}$ and $\hat{B}$ defined in \eqref{eq:scalings1} and change  to the variable $x=\rho/r_h$ in  the integral (\ref{math_I_def}). We get
\beq
\mathcal{I} = r_h^{\xi (2-q)+z-2}\int\limits_{1}^{\infty}\frac{\left(1+\hat{B}^2x^{-4\xi}\right) x ^{\xi  (2-q)+z-3}}{\left(1+\hat{B}^2x^{-4\xi}+\hat{d}^2x^{-2\xi q}\right)^{3/2}}dx\,\equiv\, r_h^{\xi  (2-q)+z-2}\,\hat{\mathcal{I}}\,.
\label{I-hat}
\eeq
Moreover, in the UV limit $r\to\infty$ the electric field $E$ can be expanded as
\beq
E =  E^{(0)}+\frac{c_E\,r^{\xi(2-q)+z-2}}{\xi(q-2)+2-z}\,+\,\ldots \ .
\eeq
We now match the expansions done in different orders. We have to compare (\ref{nh_lowfreq_diff_expansion}) and  (\ref{E_low_freq_nh_diff}).   First, we match the constant terms and get
\beq
E_{nh}=E_0+c_E\,\mathcal{I}\,\,,
\label{matching_constant_diff}
\eeq
and then the linear terms to arrive at the condition
\beq
(E_0+c_E\,\mathcal{I})\frac{i k^2 r_h^{z-3}}{\omega  \left(1+B^2r_h^{-4 \xi }+d^2r_h^{-2 \xi  q}\right)}=-c_E\frac{\left(1+B^2r_h^{-4\xi}\right) r_h^{\xi(2-q)+z-3}}{\left( 1+B^2r_h ^{-4 \xi}+d^2r_h ^{-2 \xi q}\right)^{3/2}}\,\,.
\label{matching_linear_diff}
\eeq
Imposing the Dirichlet boundary condition, $E^{(0)}=0$, (\ref{matching_linear_diff}) leads to the following   dispersion relation
\beq
\omega = -iD\,k^2\,\,,
 \label{eq:diffusion-eq}
\eeq
where $D$ is the diffusion constant, given by
\beq
D \,=\, \frac{r_h^{\xi(q-2)}\sqrt{1+\hat{B}^2+\hat{d}^2}}{1+\hat{B}^2}\,\mathcal{I}
\,=\, \frac{r_h^{z-2}\sqrt{1+\hat{B}^2+\hat{d}^2}}{1+\hat{B}^2}\,\hat{\mathcal{I}}.
\label{D_result}
\eeq
Notice that, in terms of the rescaled frequency and momentum introduced in  \eqref{eq:scalings1}, the diffusion dispersion relation can be written as:
\beq
\hat\omega\,=\,-i\,\hat D\,\hat k^2\,\,,
\eeq
where $\hat D$ is related to $D$ as:
\beq\label{hat_D_result}
\hat D\,=\,r_h^{2-z}\,D = \frac{\sqrt{1+\hat{B}^2+\hat{d}^2}}{1+\hat{B}^2}\,\hat{\mathcal{I}} \ .
 \eeq

The diffusion constant $D$ can be related to the charge susceptibility $\chi$ by means of the so-called Einstein relation, which reads:
\beq
D\,=\,\sigma\,\chi^{-{1}}\,\,,
\label{Einstein_rel}
\eeq
where $\sigma$ is the DC conductivity and $\chi$ is defined as:
\beq
\chi\,=\,{\partial\rho \over \partial\mu}\,\,,
\label{suscept_def}
\eeq
and $\rho$ is the charge density (see (\ref{density})) and $\mu$ is the chemical potential, which,  for $B\not=0$, can be written as the following integral:
\beq
\mu\,=\,d\,\int_{r_h}^{\infty}\,
{\rho^{2\xi+z-3}\over \sqrt{\rho^{2q\xi}+\rho^{2(q-2)}\,B^2\,+\,d^2}}\,d\rho\,\,.
\label{mu_B}
\eeq
From (\ref{mu_B}) it is straightforward to compute the derivative (\ref{suscept_def}) and get $\chi$. We obtain
\beq
\chi^{-1}\,=\,{\mathcal{I}\over {\cal N}}\,\,,
\eeq
where $\mathcal{I}$ is the integral (\ref{math_I_def}) and ${\cal N}$ is the normalization constant defined after (\ref{eq:DBIaction}). The DC conductivity $\sigma$ can be obtained using several techniques. In Appendix \ref{Karch-O'Bannon}, we perform this calculation for our setup, with the result:
\beq
\sigma\,=\,{\cal N}\,r_h^{\xi(q-2)}\,
 \frac{\sqrt{1+\hat{B}^2+\hat{d}^2}}{1+\hat{B}^2}\,\,.
 \eeq
It is now immediate to check that the Einstein relation (\ref{Einstein_rel}) gives the same value as our direct result (\ref{D_result}), which confirms the validity of (\ref{Einstein_rel}) for our non-relativistic background. 

The Einstein relation was postulated to hold in \cite{Kovtun:2008kx} in a holographic setting. To our knowledge \cite{Mas:2008qs} is the first work to establish its foundation concretely. Our results are the generalizations thereof.

In general, the integral $\hat{\mathcal{I}}$ cannot be evaluated analytically. However, there are two particular cases where this is not the case. These two systems are discussed in the next two subsections. 

\subsubsection{Diffusion in $2+1$ dimensions }

Let us consider the case in which $q=2$, \ie, when the field theory is $(2+1)$-dimensional. In this case the integral 
$\hat{\mathcal{I}}$ can be written in terms of the integrals $J_{\lambda_1,\lambda_2}$ defined in (\ref{J_integral_definition}) of Appendix \ref{appendix:calculations}. Actually, using (\ref{J_value}) we can write 
$\hat{\mathcal{I}}$ as:
\be
\hat{\mathcal{I}}={1\over 2-z}F\Big({3\over 2},{2-z\over 4\xi};1+{2-z\over 4\xi};-\hat d^2-\hat B^2\Big)+{\hat B^2\over 2\xi+2-z}F\Big({3\over 2},1+{2-z\over 4\xi};2+{2-z\over 4\xi};-\hat d^2-\hat B^2\Big) \ .
\ee
Moreover, using the identity 
\beq
\sqrt{1-x}\,F\Big({3\over 2}, \alpha ; \alpha+1;x\Big)\,=\,
\,F\Big(1, \alpha-{1\over 2} ; \alpha+1;x\Big)\,\,,
\label{hypergeometric_identity}
\eeq
we arrive at the following  expression of $\hat D$ for $q=2$:
\bear
&&\hat D\,=\,{1\over 1+\hat B^2}\,\Bigg[{1\over 2-z}\,
F\Big(1, {2-z-2\xi\over 4\xi} ; {2-z+4\xi\over 4\xi}
;-\hat d^2-\hat B^2\Big)\rc
&&\qquad\qquad\qquad\qquad+\,{\hat B^2\over 2-z+\xi}\,
F\Big(1, {1\over 2}+{2-z\over 4\xi} ; 2\,+\,{2-z\over 4\xi}
;-\hat d^2-\hat B^2\Big)
\Bigg]\,\,.
\label{hatD_q2}
\eear

\subsubsection{Vanishing magnetic field}

The integral $\mathcal{I}$ in (\ref{math_I_def}) can also be computed analytically when $B=0$. Using again (\ref{J_value}), we get:
\beq
D\,=\,{r_h^{z-2}\over \xi(q-2)-z+2}\,(1+r_h^{-2\xi q}\,d^2)^{{1\over 2}}\,
F\Big({3\over 2}, {\xi(q-2)-z+2\over 2\xi q}
; {\xi(3q-2)-z+2\over 2\xi q}
;-r_{h}^{-2\xi q}\,d^2\Big)\,\,.
\eeq
Notice that this expression coincides with the one obtained in section 5.2 of  \cite{Lee:2010uy} for $\xi=1$ and $q=p$. 
Moreover, this expression can be simplified by using the identity (\ref{hypergeometric_identity}), leading to the following value of the rescaled diffusion constant:
\beq
\hat D\,=\,{1\over \xi(q-2)-z+2}\,
F\Big(1, {2-z-2\xi\over 2\xi q}
; {2-z+(3q-2)\xi\over 2\xi q}
;-\hat d^2\Big)\,\,.
\label{hatD_zeroB}
\eeq
Notice that (\ref{hatD_q2}) and (\ref{hatD_zeroB}) coincide, as they should, when $q=2$ and $B=0$.

\subsubsection{Limiting behavior}
Let us return to the general case and let us study the behavior of $D$ at high and low temperature. We begin by analyzing the $T\to\infty$ limit, which corresponds to $r_h\to \infty$ and $\hat d, \hat B$ small. In this case we can neglect $\hat d, \hat B$ inside the integral (\ref{math_I_def}). If $z<2+\xi (q-2)$, we get
\beq
D \approx {r_h^{z-2}\over 
\xi(q-2)\,-\,z+2} = {1\over 
\xi(q-2)\,-\,z+2}\,\Big[{p\xi+z\over 4\pi}\Big]^{{2-z\over z}}\,T^{{z-2\over z}}\,\,,
\qquad\qquad
(T\to \infty) \ .
\eeq
This behavior matches the one found in \cite{Tong:2012nf} by applying general arguments.  We observe
that the large $T$  behavior changes qualitatively as $z$ is increased and passes through $z=2$. 

Let us next consider the low $T$ limit of $D$. In this case  we can substitute $r_h=0$ in the integral 
${\mathcal I}$ and the behavior is determined by the prefactor in (\ref{D_result}).  If $B$ is kept small enough 
($\hat B\ll 1$), we can neglect the contribution of the $B$ field to the prefactor and we get that, for small $T$, in the hydrodynamic regime, $D$ behaves as:
\beq
D\sim r_h^{-2\xi}\,\,.
\eeq
Therefore, for small $T$ the diffusion constant  $D$ behaves as:
\beq
D\sim T^{-{2\xi\over z}}\sim T^{-{2\over z}(1-{\theta\over p})}\,\,,
\qquad\qquad
(T\to 0)\,\,.
\eeq

\subsubsection{Comparison of analytic results to numerical results}

\begin{figure}[ht]
\center
 \includegraphics[width=0.40\textwidth]{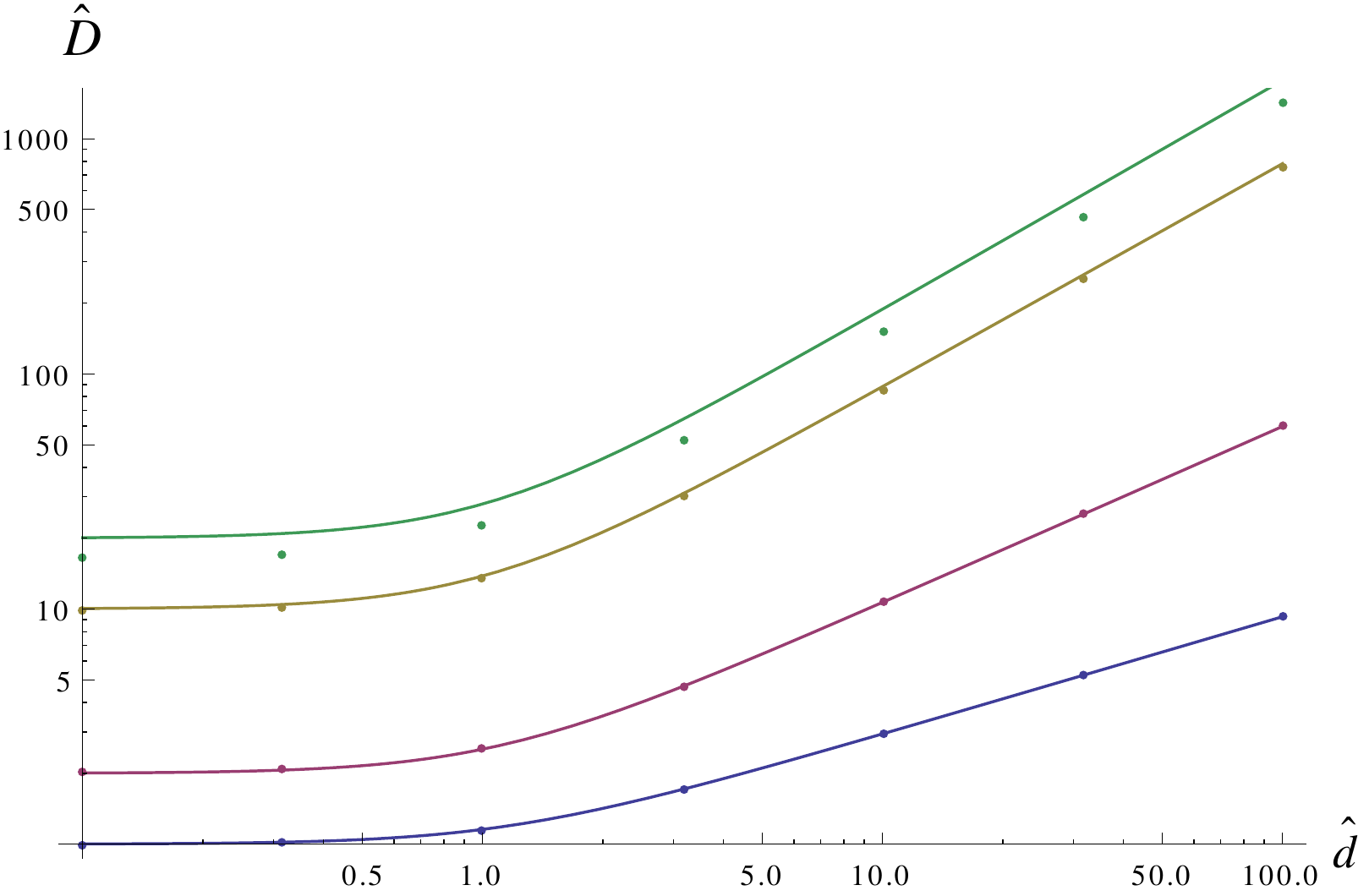}
 \qquad\qquad
  \includegraphics[width=0.40\textwidth]{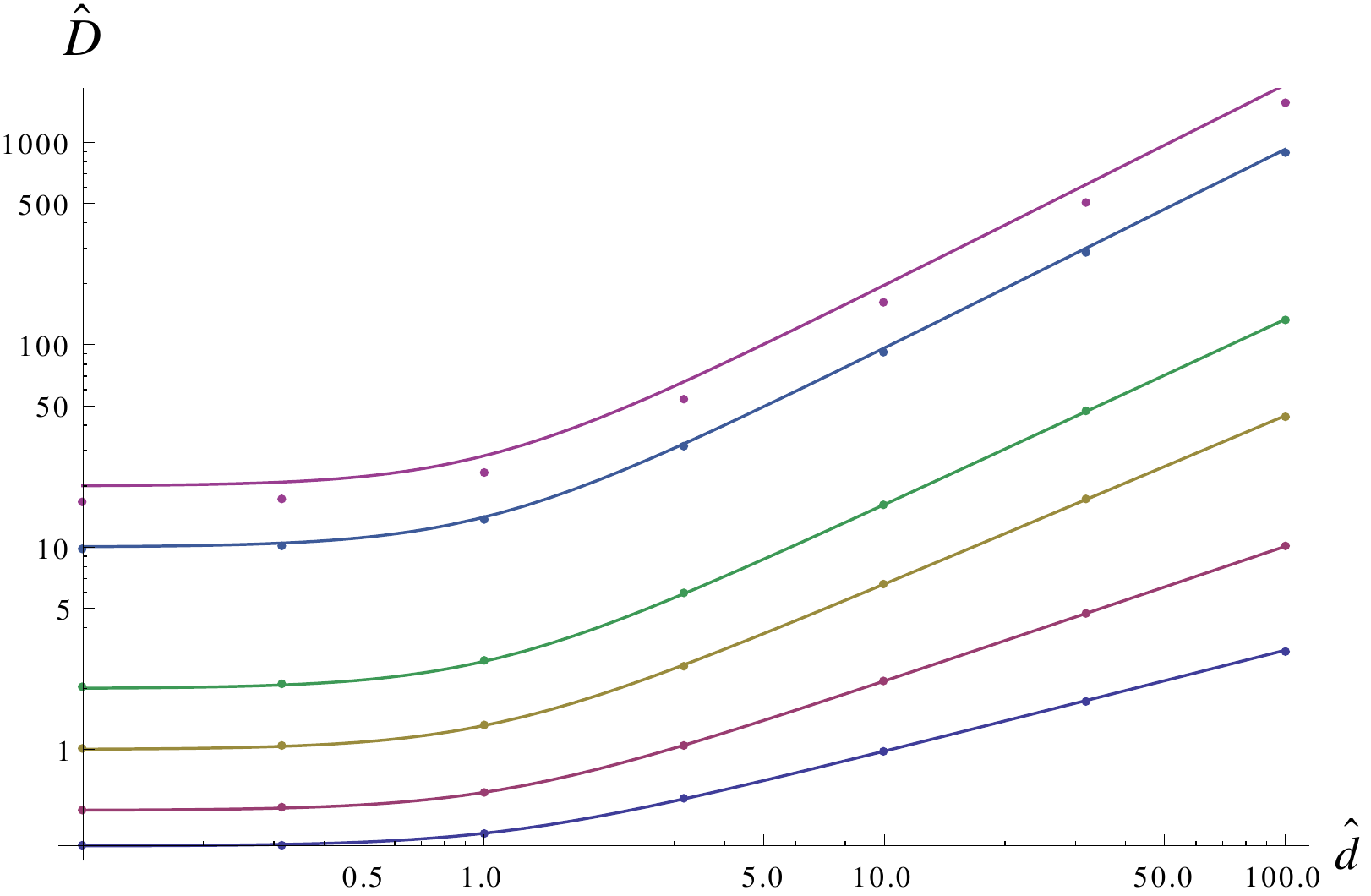}
  \caption{We depict the diffusion constant against the charge density in the absence of the magnetic field $\hat B=0$ for various cases. The numerical data is represented as points, whereas the continuous curves stem from analytic results (\ref{hatD_zeroB}). Left: We set $q=2$ and $\xi=1$. The different curves correspond to $z=1,1.5,1.9,1.95$ (bottom-up). Right: We set $q=3$ and $\xi=2$. The different curves correspond to $z=1,2,3,3.5,3.9,3.95$ (bottom-up). Notice that the plots are logarithmic.}
\label{Diffusion_zeroB}
\end{figure}

Having obtained lots of analytic results using different approximation schemes, we now wish to turn to quantifying how good they are in comparison to numerical results.
We thus compare our analytic results to those coming out of numerically solving the full fluctuation equations of motion (\ref{eq:Eeom}) and (\ref{eq:ayeom}). The numerical methods that we use are by now standard, we refer the reader to \cite{Kaminski:2009dh,Bergman:2011rf} for more details. Before direct comparisons, consider the integral $\hat{ \mathcal{I}}$ in \eqref{I-hat}. When $z$ approaches the value $\xi(q-2)+2$, it is evident that the integral becomes larger and larger, eventually diverging. It is evident that near these values, our approximation $\omega\sim k^2$ is no longer valid. 

Thus, to be more precise, we wish to find out if the diffusion mode is well-represented beyond the critical value of $z$ and what is the lower bound for 
\beq
g(q,z,\xi)\equiv\xi(q-2)-z+2 \label{eq:g}
\eeq
such that our analytical results differ from the numerical results only by a few percent. Generically then, the smaller values $g$ takes, the worse analytic results conform with the numerics.

We first consider the case without the magnetic field. Both the analytical and numerical results are represented in Fig.~\ref{Diffusion_zeroB} as functions of ${\hat d}$. We see that cleaving the two when $g\geq 0.2$ with all values of ${\hat d}$, is practically impossible. When $g\geq 0.1$ ,the analytic results are at most $5 \%$ larger than the numeric results. The difference grows slightly when considering larger values of ${\hat d}$. When $g<0.05$, the analytic result differs by more than $30 \%$ from the numerics and becomes worse as ${\hat d}$ grows. In the regime $g<0$, the diffusion mode persists  although the diffusion coefficient grows fast with decreasing $g$. Interestingly, when $g$ goes to more and more negative values, the diffusion coefficient scales as $D\propto d$ as $d\to \infty$. According to our analytic result, the diffusion constant should be linear in $d$ already at $g=0$, while numerics only support this for $g\ll 0$.

\begin{figure}[ht]
\center
 \includegraphics[width=0.40\textwidth]{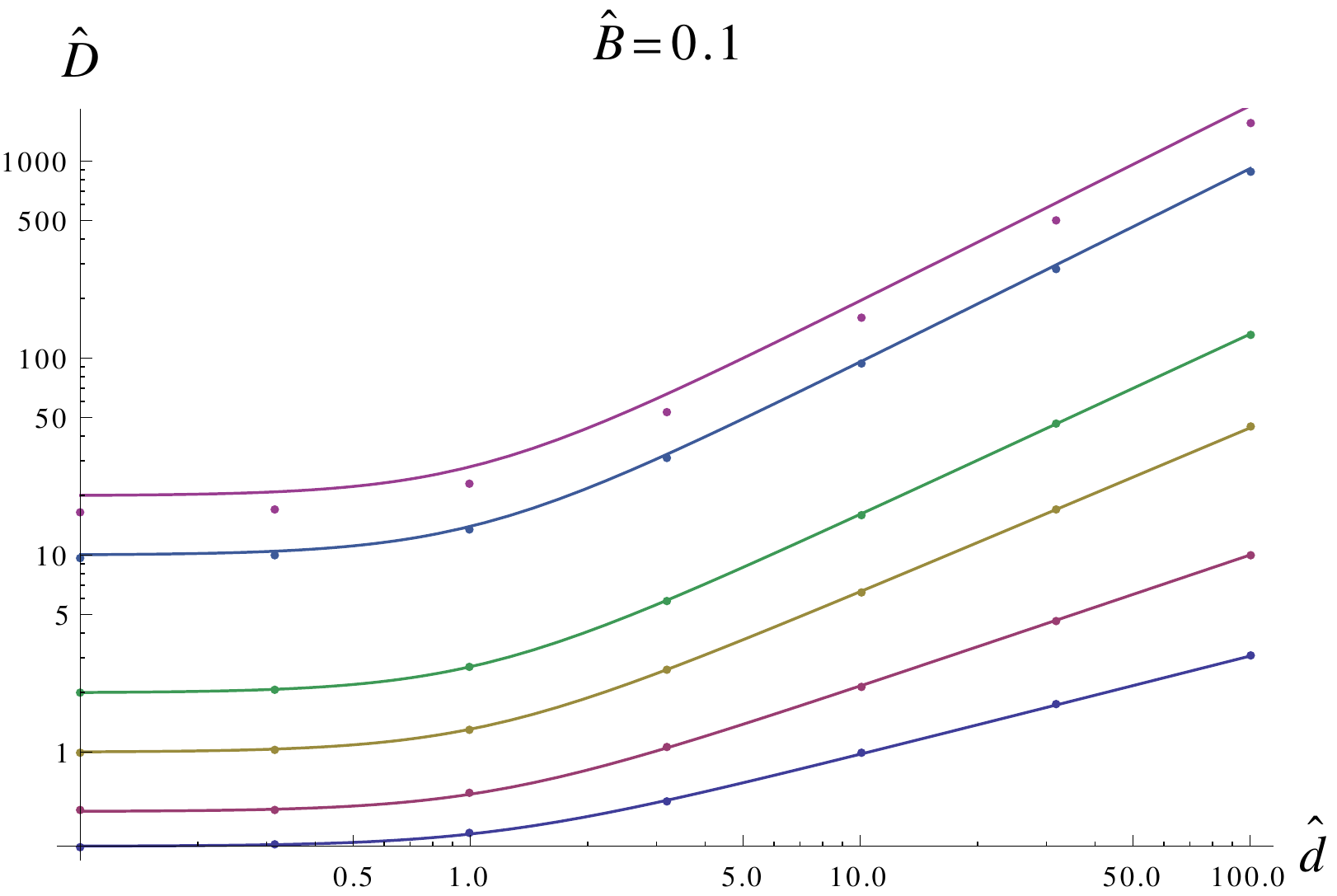}
 \qquad\qquad
  \includegraphics[width=0.40\textwidth]{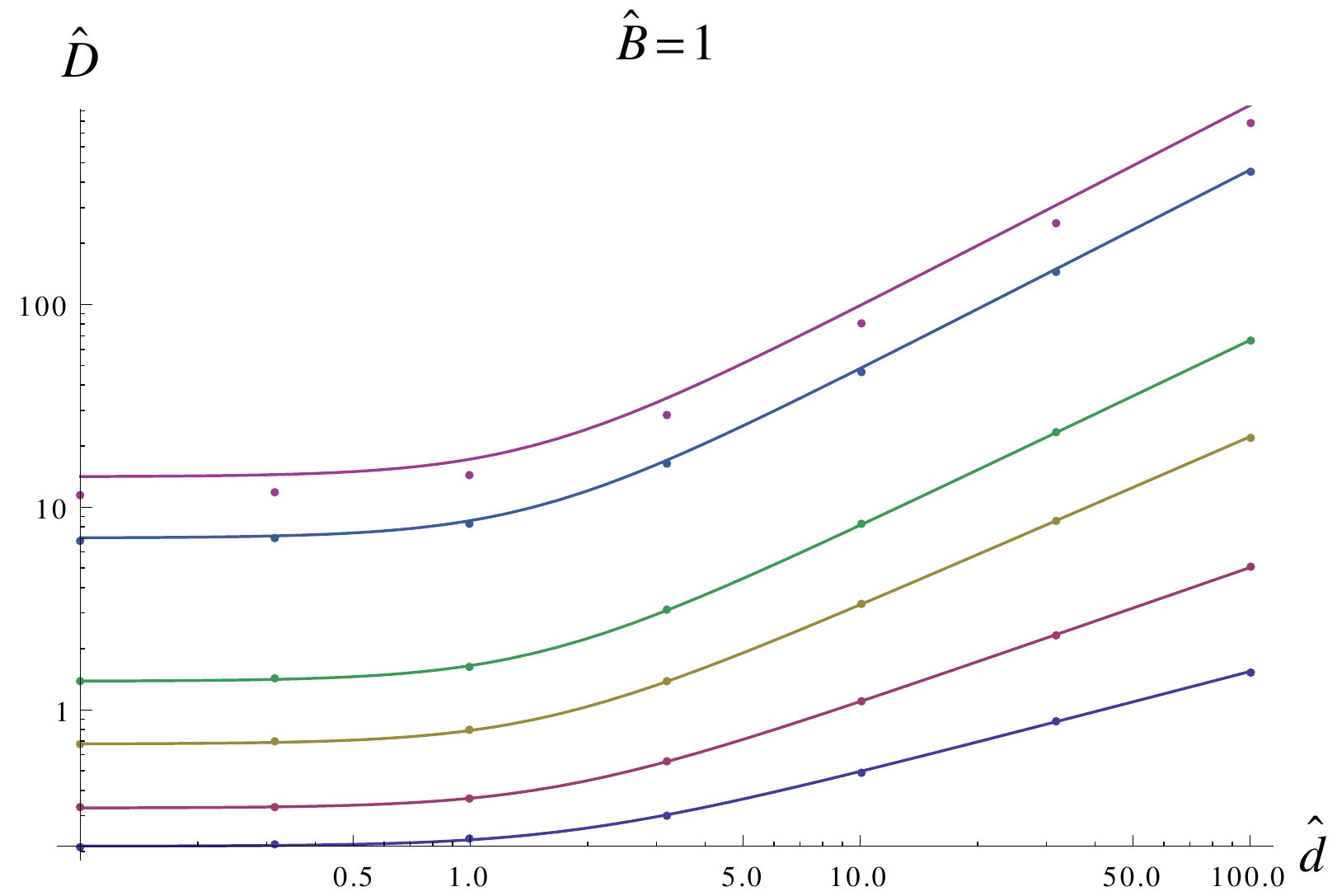}
  \\
   \includegraphics[width=0.40\textwidth]{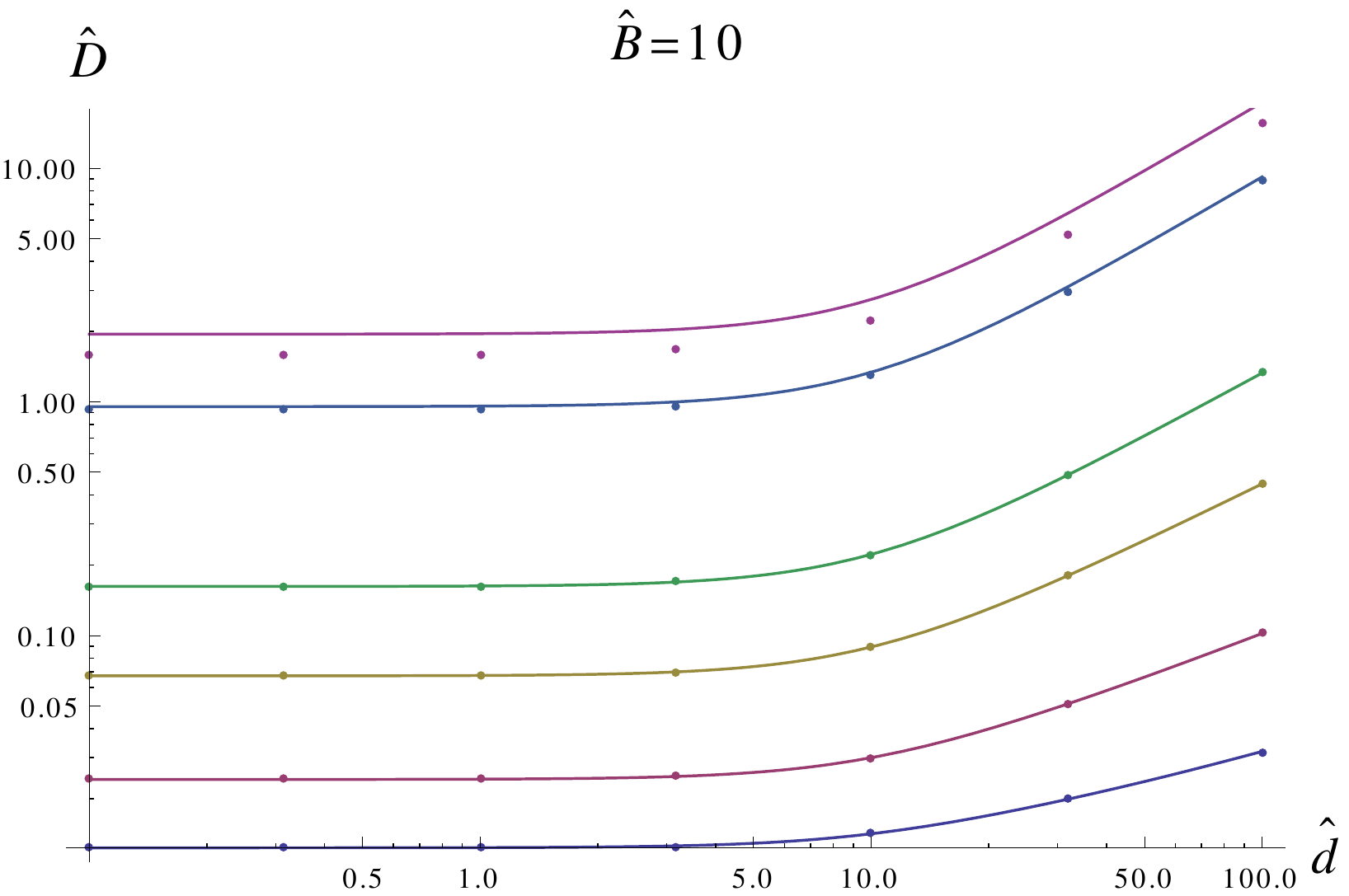}
 \qquad\qquad
  \includegraphics[width=0.40\textwidth]{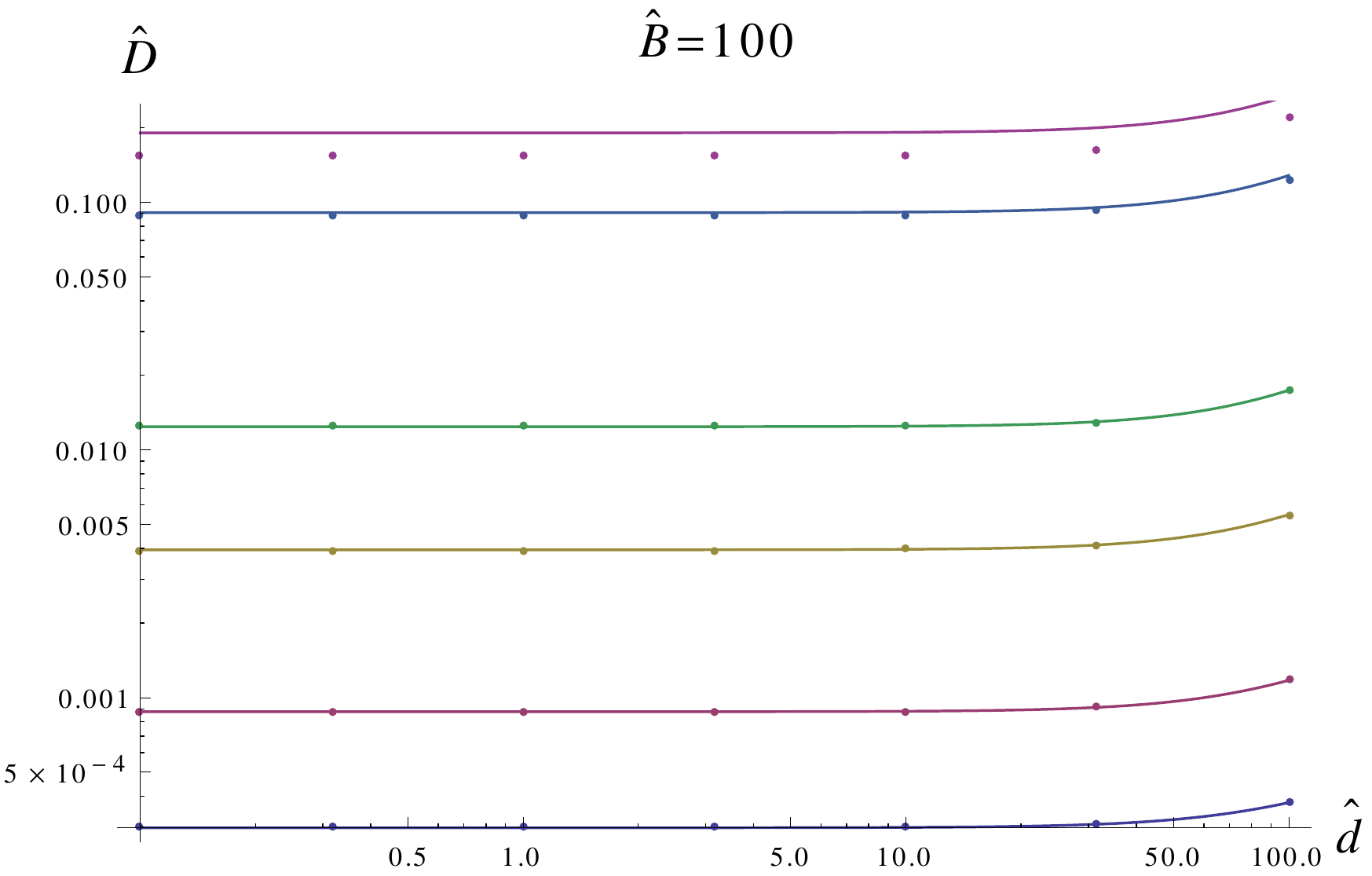}
  \caption{Comparison of the numerically computed diffusion constant $\hat D$ as a function of the charge density $\hat d$ with that of analytic prediction. Different panels focus on different magnetic field strengths as indicated in the plots. The points stand for numerical data while the curves follow from formulas (\ref{hat_D_result}). We focus on $q=3$, $\xi=2$ for all the cases. The different curves have dynamical exponent $z=1,2,3,3.5,3.9,3.95$ (bottom-up). Notice that the plots are logarithmic.}
\label{Diffusion_B}
\end{figure}

Now, let us consider the effect of the magnetic field. The analytic and numerical results are compared in Fig.~\ref{Diffusion_B} in various different cases. Again, we see that for largish values of $g$, the results agree very well.

\subsection{Zero sound}\label{sec:zerosound}

Let us now study the system in the collisionless regime. With this purpose, 
let us consider the equations for the fluctuations (\ref{eq:Eeom}) and (\ref{eq:ayeom}) at zero temperature and non-zero $B$ field. We  will assume that $B$ is small. Near the horizon  $r=0$ the equations (\ref{eq:Eeom}) and (\ref{eq:ayeom}) read:
\bear
E''\,+\,\Bigg({z+1-2\xi\over r}+{4\xi B^2\over r(r^{4\xi}+B^2)}
\Bigg)
\,E'\,+\,{\omega^2\over r^{2z+2}}\,E & = &
-4i\,B\,\xi\,\omega^2\,{r^{2\xi-z-2}\over r^{4\xi}+B^2}\,a_y\rc
a_y''\,+\,\Bigg({z+1-2\xi\over r}+{4\xi B^2\over r(r^{4\xi}+B^2)}
\Bigg)
\,a_y'\,+\,{\omega^2\over r^{2z+2}}\,a_y & = & 
4i\,B\,\xi\,{r^{2\xi-z-2}\over r^{4\xi}+B^2}\,E\,\,.
\eear
Let us define an operator $\hat {\cal O}$ as the one that acts on any function $F(r)$ as follows:
\beq
\hat {\cal O}\,F\,\equiv\,F''\,+\,\Bigg({z+1-2\xi\over r}+{4\xi B^2\over r(r^{4\xi}+B^2)}
\Bigg) \,F'\,+\,{\omega^2\over r^{2z+2}}\,F \ .
\eeq
Then, the system of coupled equations can be written as:
\beq
\hat {\cal O}\,E\,=\,-4i\,B\,\xi\,\omega^2\,{r^{2\xi-z-2}\over r^{4\xi}+B^2}\,a_y\,\,,
\qquad\qquad
\hat {\cal O}\,a_y\,=\,4i\,B\,\xi\,{r^{2\xi-z-2}\over r^{4\xi}+B^2}\,E\,\,.
\eeq
These equations can be decoupled. Let us define the functions 
\beq
y_{\pm}(r)\,=\,{E\over i\omega}\,\pm\,a_y\,\,.
\eeq
The equations for $y_{\pm}(r)$ are:
\beq
\Bigg(\hat {\cal O}\pm 4\,B\,\xi\omega\,{r^{2\xi-z-2}\over r^{4\xi}+B^2}\Bigg)\,y_{\pm}\,=\,0 \ .
\eeq
Let us study these equations when $B$ is small. Neglecting the terms that are quadratic in $B$, we get:
\beq
y_{\pm}''\,+\,{z+1-2\xi\over r}
\,y_{\pm}'\,+\,\Bigg({\omega^2\over r^{2z+2}}\pm {4\xi \omega B \over r^{2\xi+z+2}}
\Bigg)\,y_{\pm}\,=\,0 \ .
\eeq
We now solve this equation in powers of $B$. As the equation for $y_+$ is obtained from the equation of $y_-$ by changing $B$ by $-B$, we can write (at first order in $B$):
\beq
y_{\pm}(r)=y_0(r)\pm B\, y_1(r)\,\,.
\eeq
The equations of $y_0$ and $y_1$ do not depend on $B$ and are given by:
\bear
&&y_{0}''\,+\,{z+1-2\xi\over r}
\,y_{0}'\,+\,{\omega^2\over r^{2z+2}}
\,y_{0}\,=\,0\rc
&&y_{1}''\,+\,{z+1-2\xi\over r}
\,y_{1}'\,+\,{\omega^2\over r^{2z+2}}
\,y_{1}\,=\,-{4\,\xi\, \omega  \over r^{2\xi+z+2}}\,y_0\,\,.
\eear
The solution for $y_0$ with infalling boundary conditions can be  written in terms of a Hankel function as:
\beq
y_0(r)\,=\,c_+\, r^{\xi-{z\over 2}}\,H^{(1)}_{{1\over 2}-{\xi\over z}}\,
\Big({\omega\over z\,r^{z}}\Big)\,\,.
\eeq
Moreover, if we write $y_1=c_+\,y$, we get the following equation for $y$:
\beq
y''\,+\,{z+1-2\xi\over r}
\,y'\,+\,{\omega^2\over r^{2z+2}}
\,y\,=\,-{4\,\xi\, \omega  \over r^{\xi+{3z\over 2}+2}}\,
H^{(1)}_{{1\over 2}-{\xi\over z}}\,
\Big({\omega\over z\,r^{z}}\Big)\,\,.
\label{inhomo_y_eq}
\eeq
This equation is solved in appendix \ref{wronskian} by using the Wronskian method, with the result:
\beq
y(r)\,=\,r^{-\xi-{z\over 2}}\,
H_{-{\xi\over z}-{1\over 2}}^{(1)}\Big({\omega\over z r^z}\Big)\,\,.
\eeq

Let us now write the complete near-horizon solution. First we define the functions
$z_1(r)$ and $z_2(r)$ as:
\beq
z_1(r)\equiv r^{\xi-{z\over 2}}\,H^{(1)}_{{1\over 2}-{\xi\over z}}\,
\Big({\omega\over z\,r^{z}}\Big)\,\,,
\qquad\qquad
z_2(r)\equiv r^{-\xi-{z\over 2}}\,
H_{-{\xi\over z}-{1\over 2}}^{(1)}\Big({\omega\over z\, r^z}\Big)\,\,.
\eeq
Then, $y_{\pm}(r)$ are given by:
\beq
y_{+}(r)\,=\,c_+\,z_1(r)\,+\,c_+\,B\,z_2(r)\,\,,
\qquad\qquad
y_{-}(r)\,=\,c_-\,z_1(r)\,-\,c_-\,B\,z_2(r)\,\,,
\eeq
where, to obtain $y_{-}$ we changed $B\to -B$ and $c_+\to c_-$. Let us now redefine these constants as follows
\beq
c_1\,=\,{i\omega\over 2}\,(c_++c_-)\,\,,
\qquad\qquad\qquad\qquad
c_2\,=\,{i\omega\over 2}\,(c_+-c_-)\,\,.
\eeq
Then, $E$ and $a_y$ can be written in matrix form as:
\beq
\begin{pmatrix}
E\\ \\ a_y
\end{pmatrix}
\,=\,
\begin{pmatrix}
  z_1(r) &&& B\,z_2(r)\\
  {} & {} \\
 - {iB\over \omega}\,z_2(r) &&& -{i\over \omega}\,z_1(r)
 \end{pmatrix}\,
 \begin{pmatrix}
c_1\\ \\c_2
\end{pmatrix}\,\,.
\eeq
Let us now consider this solution at low frequency ($\omega r^{-z}\ll 1$).
When the index $\nu$ is not integer (\ie, for  $z\not= 2\xi$) the Hankel function has the following expansion near the origin:
\beq
H_{\nu}^{(1)}(\alpha x)= -{2^{\nu}\,\Gamma(\nu)\over \pi \alpha^{\nu}}\,i\,
\Big[{1\over x^{\nu}}\,+\,{\pi\over \Gamma(\nu)\,\Gamma(\nu+1)}\,
\Big({\alpha\over 2}\Big)^{2\nu}\,
\Big(i\,-\,\cot (\pi\nu)\Big)\,x^{\nu}\,+\,\ldots\Big]\,\,,
\eeq
Using this expression, we  expand $z_1(r)$ and $z_2(r)$ for low $\omega$ as:
\beq
z_1(r)\approx -{2^{\nu}\,\Gamma(\nu)\over \pi \alpha^{\nu}}\,i\,
\Big[1\,+\,c\,\omega^{1-{2\xi\over z}}\,r^{2\xi-z}\,\Big]\,\,,
\qquad
z_2(r)\approx -{2^{\nu}\,\Gamma(\nu)\over \pi \alpha^{\nu}}\,i\,
\Big[ c (z-2\xi)\,\omega^{-{2\xi\over z}}\Big]\,\,,
\eeq
where $\nu$ is the index written in (\ref{nu_dz}), $\alpha=\omega/z$ and $c$ is the constant 
\beq
c\,=\,{\pi\over z-2\xi}\,
{(2z)^{{2\xi\over z}}\over \Gamma\Big({1\over 2}-{\xi\over z}\Big)^2}\,
\Big[i-\tan\Big({\pi\xi\over z}\Big)\Big]\,\,.
\label{c_def}
\eeq
By absorbing the common factor in $z_1$ and $z_2$, we have:
\beq
\begin{pmatrix}
E\\ \\ a_y
\end{pmatrix}
\,\approx\,
\begin{pmatrix}
  1+c\,
 \omega^{1-{2\xi\over z}} r^{2\xi-z}&&& B\,c(z-2\xi)\,\omega^{-{2\xi\over z}}
 \\
  {} & {} \\
  -iB\,c(z-2\xi)\,\omega^{-1-{2\xi\over z}}
  &&& -{i\over \omega}\,
  \big(1+c\,\omega^{1-{2\xi\over z}} r^{2\xi-z} \big)
 \end{pmatrix}\,
 \begin{pmatrix}
c_1\\ \\c_2
\end{pmatrix}\,\,.
\label{E_ay_nh_low}
\eeq
\subsubsection{Matching}

Let us now obtain the values of $E$ and $a_y$ when the near-horizon and low frequency limits are taken in the opposite order.  We thus consider $\omega$ and $k$ being small and of the same order. It is easy to see, that in this limit, the equation of $E$ decouples from that of $a_y$ and that one can neglect the terms without derivatives in (\ref{eq:Eeom}).  Moreover, we will assume that $B$ is small and, therefore, we will just take $B=0$ in the remaining terms in (\ref{eq:Eeom}). After these approximations, the equation for $E$ at low frequency becomes:
\beq
E''\,+\,\partial_r\,\log\Bigg[
{r^{z+1-2\xi}(r^{2\xi q}+d^2)^{{3\over 2}}\over
(\omega^2-r^{2z-2}\,k^2)\,r^{2\xi q}\,+\,\omega^2\,d^2}\Bigg]
\,E'\,=\,0\,\,.
\eeq
This equation can be readily integrated:
\beq
E'\,=\,c_E\,
{(\omega^2-r^{2z-2}\,k^2)\,r^{2\xi q}\,+\,\omega^2\,d^2\over
r^{z+1-2\xi}(r^{2\xi q}+d^2)^{{3\over 2}}}\,\,,
\eeq
with $c_E$ being a constant of integration. A second integration yields:
\beq
E(r)\,=\,E^{(0)}\,-\,c_E\,\big[\,\omega^2\, I(r)\,-\,k^2\,J(r)\big]\,\,, \label{E-sol-zs-small-k}
\eeq
where $E^{(0)}=E(r\to\infty)$ and $I(r)$ and $J(r)$ are defined in terms of the integrals 
(\ref{I_lambda12_def})  and (\ref{J_integral_definition}) of Appendix \ref{appendix:calculations}:
\beq
 I(r) = I_{2\xi-z-1\,, \,2\xi q}(r)\,\,,
 \qquad\qquad
 J(r) = J_{2\xi q+2\xi+z-3, 2\xi q}(r)\,\,.
 \label{I_J_def}
\eeq
From (\ref{I_lambda12_value}) and (\ref{J_value})  we get:
\bear
{ I}(r) & = & {r^{\xi(2-q)-z}\over z+\xi(q-2)}\,
F\Big({1\over 2},{z+\xi(q-2)\over 2\xi q};
{z+\xi(3q-2)\over 2\xi q};-r^{-2\xi q}\,d^2\Big)\rc
J(r) &= &
{r^{\xi(2-q)+z-2}\over \xi(q-2)-z+2}\,
F\Big({3\over 2}, {\xi(q-2)-z+2\over 2\xi q}
; {\xi(3q-2)-z+2\over 2\xi q};-r^{-2\xi q}\,d^2\Big)\,\,.\qquad\qquad
\eear
Let us now expand ${ I}(r)$ and ${ J}(r)$ near the horizon $r=0$. From (\ref{I_lambda12_expansion}) we have:
\beq
 I(r)\,=\,I_0+{r^{2\xi-z}\over z-2\xi}\,d^{-1}\,+\,\ldots\,\,,
 \label{I_nh}
\eeq
with
\beq
I_0\,=\,{1\over 2\xi q}\,B\Big({2\xi-z\over 2\xi q}\,,\,{z+\xi(q-2)\over 2\xi q}\Big)\,
d^{{2\xi-z\over \xi q}-1}\,\,.
\label{I_0}
\eeq
Moreover, at the order we are working we can take $ J(r)$ as its value at $r=0$:
\beq
 J(r)\approx J_0\,=\,{2\xi+z-2\over 2\xi^2\,q^2}\,
B\Big({2\xi+z-2\over 2\xi q}\,,\,{1\over 2}\,-\,{2\xi+z-2\over 2\xi q}\Big)\,
d^{{2\xi+z-2\over \xi q}-1}\,\,.
\eeq
Therefore, near $r=0$, we have:
\beq
E(r)\approx -{c_E\over z-2\xi}\,{\omega^2\over d}\,r^{2\xi-z}\,+\,E^{(0)}\,-\,
c_E\,\big(\omega^2\,I_0-k^2\,J_0\big)\,\,.
\label{E_low_nh}
\eeq

Let us now consider the fluctuations of $a_y$. At low frequency and small $B$, (\ref{eq:ayeom}) becomes:
\beq
a_y''\,+\,\partial_r\,\Big[r^{z+1-2\xi}\,(r^{2\xi q}\,+\,d^2)^{{1\over 2}}\Big]\,a_y'\,=\,0\,\,.
\eeq
This equation can be integrated once as:
\beq
a_y'\,=\,{c_y\over r^{z+1-2\xi}\,(r^{2\xi q}\,+\,d^2)^{{1\over 2}}}\,\,,
\eeq
where $c_y$ is an integration constant. An additional integration yields:
\beq
a_y(r)\,=\,a_y^{(0)}\,-\,c_y\,\int_{r}^{\infty}\,
{d\rho \over \rho^{z+1-2\xi}\,(\rho^{2\xi q}+d^2)^{{1\over 2}}}\,\,. \label{ay-sol-zs-small-k}
\eeq
Clearly, $a_y^{(0)}=a_y(r\to \infty)$. Moreover, $a_y(r)$ can be written in terms of the integral ${ I}(r)$ defined in (\ref{I_J_def}). We get:
\beq
a_y(r)\,=\,a_y^{(0)}\,-\,c_y\, {\cal I}(r)\,\,.
\eeq
The expansion of $a_y(r)$ near the horizon $r=0$ can be readily obtained from (\ref{I_nh}). We get:
\beq
a_y(r)\,=\,a_y^{(0)}\,-\,c_y\,I_0\,-\,{c_y\over (z-2\xi)d}\,r^{2\xi-z}\,\,,
\label{ay_low_nh}
\eeq
where $I_0$ has been defined in (\ref{I_0}). 

Let us now compare (\ref{E_low_nh}) and (\ref{ay_low_nh}) to (\ref{E_ay_nh_low}). From the terms depending on $r$ we get:
\beq
c_1\,=\,-{c_E\over (z-2\xi)c}\,{\omega^{1+{2\xi\over z}}\over d}\,\,,
\qquad\quad
c_2\,=\,-i\,{c_y\over (z-2\xi)c}\,{\omega^{{2\xi\over z}}\over d}\,\,.
\eeq
Using these results we get the following matrix relation from the comparison of the constant terms:
\beq
\begin{pmatrix}
E^{(0)}\\ \\ a_y^{(0)}
\end{pmatrix}
\,=\,
\begin{pmatrix}
\omega^2\,I_0\,-\,k^2\,J_0\,-\,{\omega^{1+{2\xi\over z}}\over (z-2\xi)\,c\,d}
 &&&-i{B\over d}\\
  {} & {} \\
  i{B\over d} &&& 
 I_0-{\omega^{1+{2\xi\over z}}\over (z-2\xi)\,c\,d}
  \end{pmatrix}\,
 \begin{pmatrix}
c_E\\ \\c_y
\end{pmatrix}\,\,.
\label{E0_ay0_B}
\eeq
The non-trivial solution in which the sources $E^{(0)}$ and $a_y^{(0)}$ vanish only exists when the determinant of matrix in (\ref{E0_ay0_B}) is zero. This condition is equivalent to the equation:
\beq
\Big(\omega^2\,-\,{J_0\over I_0}\,k^2\,\,-\,{\omega^{1+{2\xi\over z}}\over (z-2\xi)\,c\,d I_0}\Big)\,
\Big(1-{\omega^{{2\xi\over z}-1}\over (z-2\xi)\,c\,d I_0}\Big)\,=\,\Big({B\over d\,I_0}\Big)^2\,\,,
\label{zero_disp_general}
\eeq
which determines the dispersion relation of the zero sound. We will analyze this in great detail by starting with the vanishing magnetic field case.

\subsubsection{Vanishing magnetic field}

\begin{figure}[ht]
\center
 \includegraphics[width=0.40\textwidth]{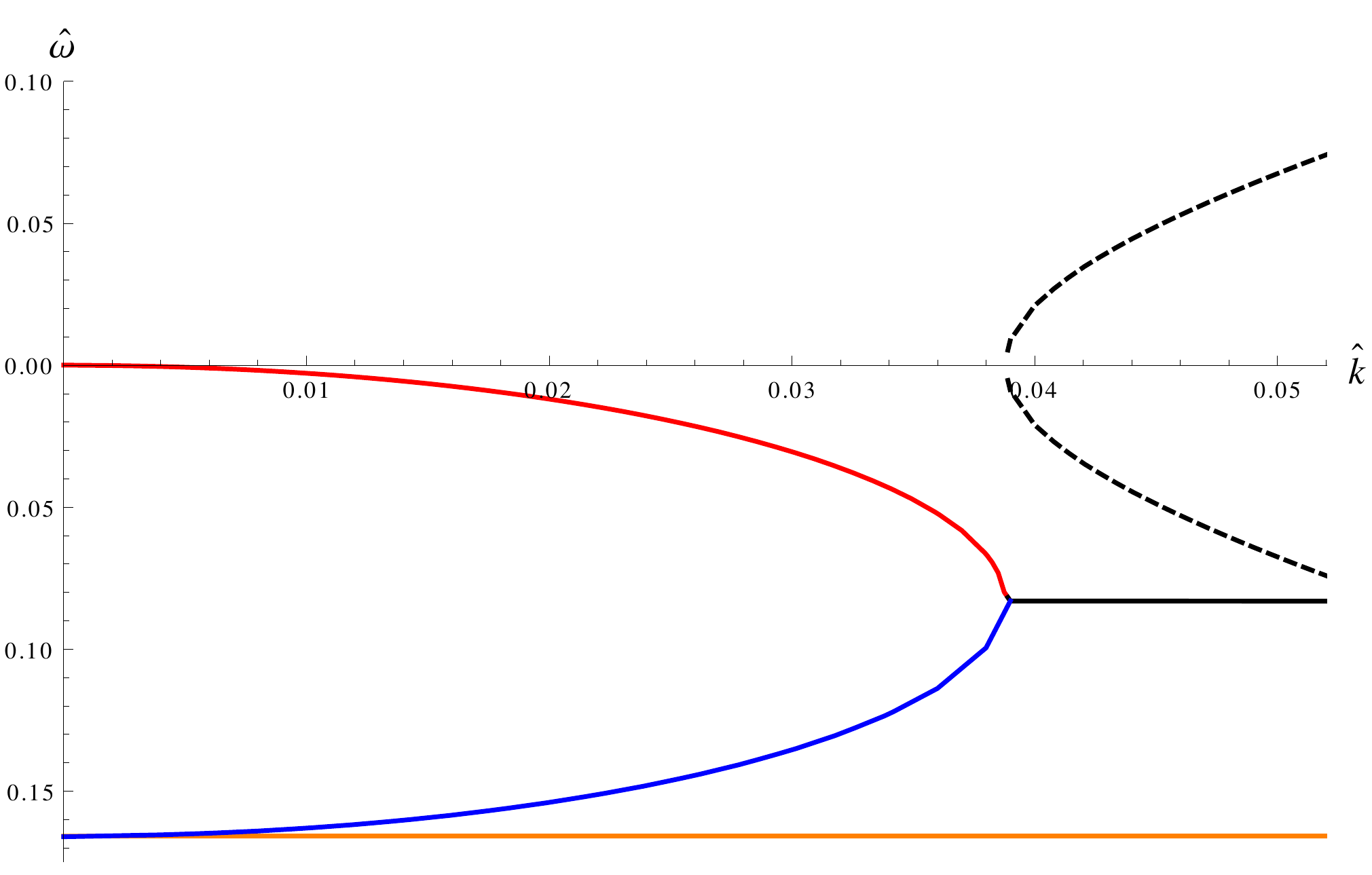}
   \caption{We present a typical dispersion relation at low energy at vanishing magnetic field strength $\hat B=0$. The parameters here are chosen such that $q=3$, $z=1.5$, $\xi = 1$, and $\hat d= 2000$. The continuous curves denote imaginary parts of the modes, whereas dashed curves represent the corresponding real parts, if non-vanishing. The red curve starting from the origin is the diffusion mode which merges at $\hat k\sim 0.04$ with another purely imaginary mode (blue curve), that can be traced to the longitudinal excitation mode (can be identified $\hat k=0$). At $\hat k=0$ there is degeneracy for the longitudinal and transverse (orange curve) gauge field fluctuations. The merging point $\hat k\sim 0.04$ defines the transition point from the hydrodynamical regime to the collisionless regime. Beyond this point, the lowest excitation mode is the zero sound (black curves).}
\label{Dispersion_relation}
\end{figure}

Let us consider in detail the case $B=0$. By imposing the Dirichlet boundary condition 
 $E^{(0)}=0$ in (\ref{E0_ay0_B}), we get the following dispersion relation:
\beq\label{zero_disp_general_B0}
\omega^2\,I_0-k^2\,J_0\,=\,
{\omega^{1+{2\xi\over z}}\over (z-2\xi)\,c\,d}\,\,.
\eeq
At leading order we can neglect the right-hand side of the equation, which leads to:
\beq\label{eq:omega0B0}
\omega\,=\,\pm\sqrt{{J_0\over I_0}}\,\,k\,\,.
\eeq
Thus, the speed of zero sound is:
\beq
c_s^2\,=\,{J_0\over I_0}={2\xi+z-2\over \xi\,q}\,d^{{2(z-1)\over \xi q}}\,
{B\Big({2\xi+z-2\over 2\xi q}\,,\,{1\over 2}\,-\,{2\xi+z-2\over 2\xi q}\Big)\over
B\Big({2\xi-z\over 2\xi q}\,,\,{z+\xi(q-2)\over 2\xi q}\Big)}\,\,,
\label{c_s}
\eeq
in agreement with the result found in \cite{Dey:2013vja}. Interestingly, this differs from the speed of first sound \eqref{eq:speed-1st-sound} only by the power of $d$ and the Euler Beta-functions. Also, it exactly reduces to the speed of first sound when $z=1$. Let us next take into account the next order contribution and write the more general expression in (\ref{zero_disp_general_B0}) in such a way that a comparison to the results of \cite{Dey:2013vja} is transparent. First of all, we recast the constant $c$ defined in (\ref{c_def}) as:
\beq
c\,=\,{(2z)^{{2\xi\over z}-1}\over \pi}\,
{\Gamma\Big({\xi\over z}+{1\over 2}\Big)\,\Gamma\Big({\xi\over z}-{1\over 2}\Big)\over
i+\tan\Big({\pi \xi\over z}\Big)}\,\,.
\eeq
Then, we can verify that:
\beq
(z-2\xi)\,c\,J_0\,d\,=\,-{\alpha_1\over \alpha_3}\,\,,
\eeq
where $\alpha_1$ and $\alpha_3$ are the constants defined in \cite{Dey:2013vja}:
\bear
\alpha_1 & = & {(2\xi-z)(2\xi+z-2)\,(2z)^{{2\xi\over z}-1}\over 
2\xi^2\,q^2\,\pi d^{{2-2\xi-z\over \xi q}}}\,\,
\Gamma\Big({\xi\over z}+{1\over 2}\Big)\,\Gamma\Big({\xi\over z}-{1\over 2}\Big)\rc
&&\qquad\qquad\qquad\qquad\qquad
\times B\Big({2\xi+z-2\over 2\xi q}\,,\,{1\over 2}\,-\,{2\xi+z-2\over 2\xi q}\Big)\rc
\alpha_3 & = &i+\tan\Big({\pi \xi\over z}\Big)\,\,.
\eear
The dispersion relation can now be written as in \cite{Dey:2013vja}:
\beq
k^2\,-\,{\omega^2\over c_s^2}\,-\,{\alpha_3\over \alpha_1}\,\omega^{1+{2\xi\over z}}\,=\,0\,\,.
\eeq

The next-to-leading order contribution to gives an imaginary part for $\omega$. Introducing $\delta\omega=\omega-c_s k$, we can solve to linear order
\bear
&{\rm{Im}} \,\delta \omega&= -\frac{c_s^{2+\frac{2\xi}{z}}}{2\alpha_1}k^{\frac{2\xi}{z}}\\
\!\!\!\!\! &=& \!\!\!\!\!\!\!\!-\frac{(2z^2)^{-\frac{\xi}{z}}d^{\frac{z^2-2\xi}{q z \xi}}\pi q \xi }{B\left(\frac{2\xi-z}{2\xi q},\frac{z+(q-2)\xi}{2 \xi q}\right)\Gamma\left(\frac{1}{2}+\frac{\xi}{z}\right)^2 }\left( \frac{\Gamma \left(\frac{2-z+(q-2) \xi}{2 q \xi }\right) \Gamma \left(\frac{z+2 \xi -2}{2 q \xi }+1\right)}{\Gamma
   \left(\frac{2\xi-z}{2 q \xi}\right) \Gamma \left(\frac{z+(q-2) \xi }{2 q \xi }\right)}  \right)^{\frac{\xi}{z}} k^{\frac{2\xi}{z}} \ .
\eear

\subsubsection{Non-zero magnetic field}

Let us continue solving (\ref{zero_disp_general}) at non-zero $B\ne 0$ but at leading order in frequency and momentum. In this case, we can write
\beq\label{eq:omatBne0withk}
\omega^2\,=\,c_s^2\,k^2\,+\,\omega_0^2\,\,,
\eeq
where  $c_s$ is the speed of zero sound written in (\ref{c_s}) and the gap $\omega_0$ is 
\beq
\omega_0\,=\,{B\over d\,I_0}\,\,.
\eeq
More explicitly, $\omega_0$ can be written as:
\beq\label{eq:omatBne0}
\omega_0\,=\,{2\,\xi\,q\,B\,d^{{z-2\xi\over \xi q}}\over 
B\Big({2\xi-z\over 2\xi q}\,,\,{z+\xi(q-2)\over 2\xi q}\Big)
}\,\,.
\eeq

\begin{figure}[ht]
\center
 \includegraphics[width=0.40\textwidth]{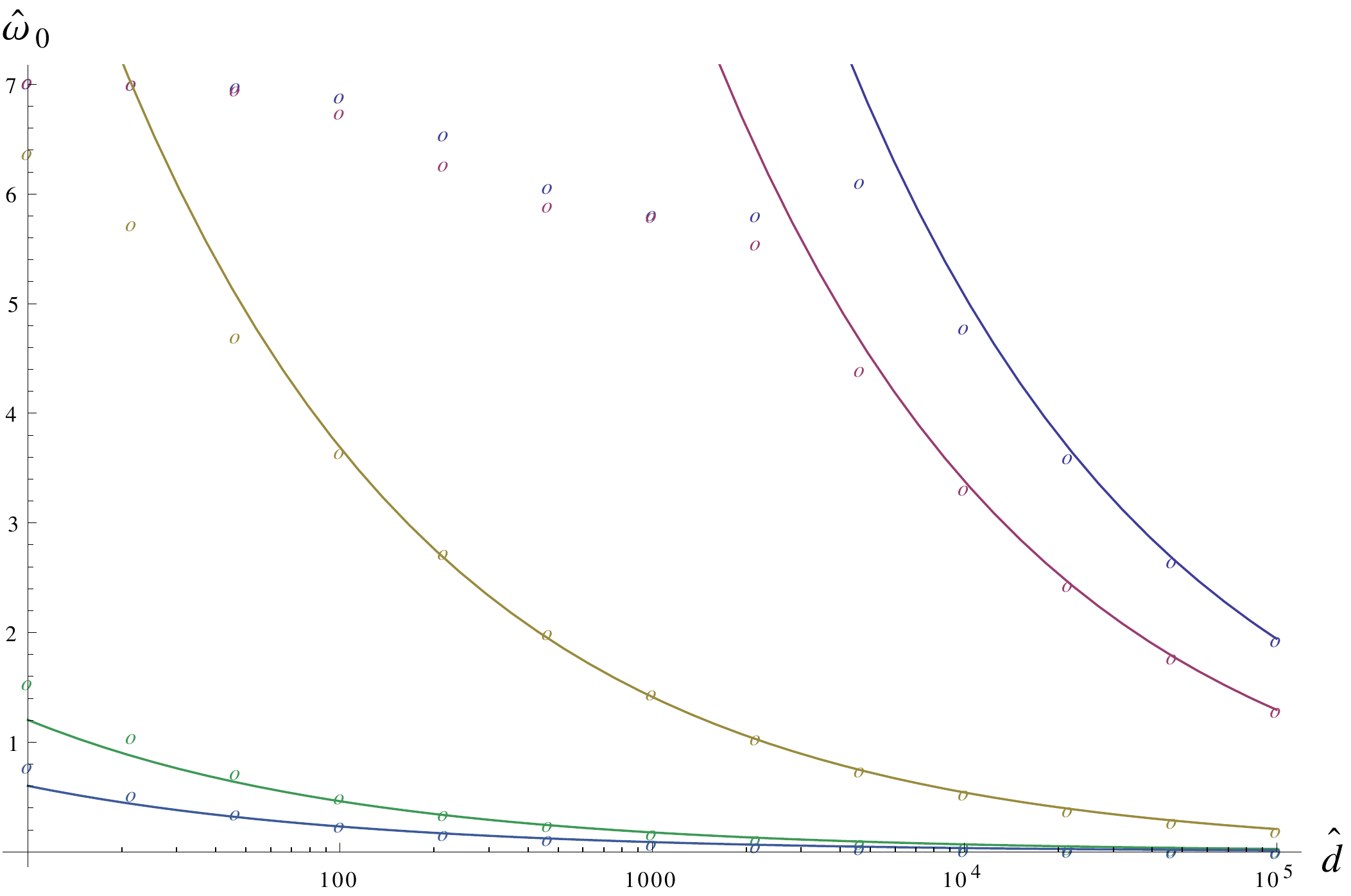}
 \qquad\qquad
  \includegraphics[width=0.40\textwidth]{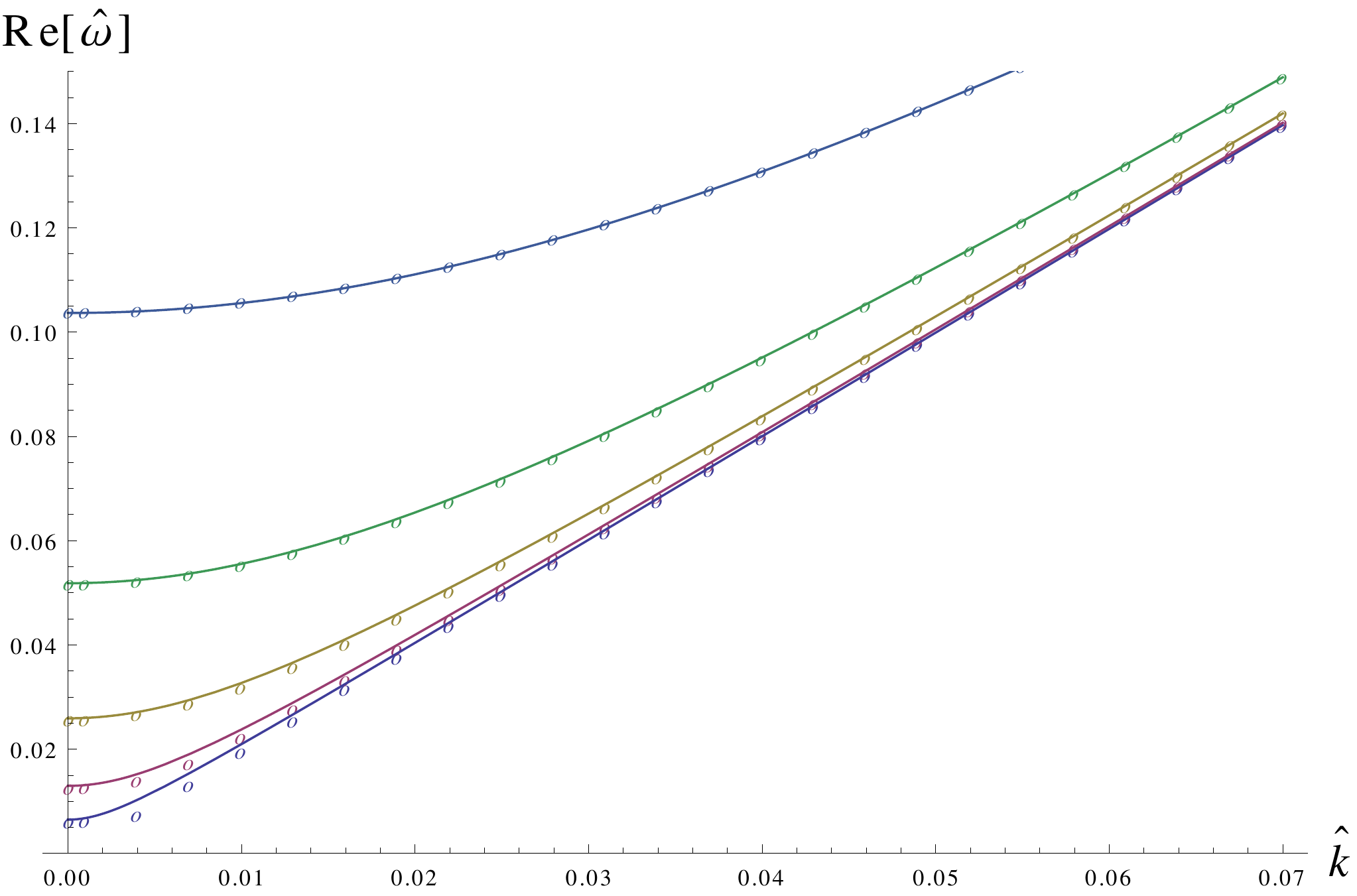}
\caption{Left: We present a comparison of numerical computation of the gap in the zero sound mode to our analytical result (\ref{eq:omatBne0}) by varying $\hat d$. The parameters are $q =3$, $\xi = 2$, and $z = 1.5$. The different curves correspond to different values of the magnetic field:  $\hat B=1, 2, 16, 100, 150$ (bottom-up). Right: We show a comparison of  the numerical and analytical results (\ref{eq:omatBne0withk}) for the  real part of the zero sound mode as a function of $\hat k$. The parameters are: $q=3$, $z=1.5$, $\xi = 2$, and $\hat d = 10^5$. The values used for the magnetic field are $\hat B=0.5$, $1$, $ 2$, $4$, and $8$ (bottom-up).}
\label{zero_sound_gap_B}
\end{figure}

We notice that the mass gap is linear in the magnetic field strength, in accordance with the Kohn's theorem. At finite temperature, we expect that there is a critical magnetic field above which the zero sound will acquire a mass \cite{Jokela:2012vn} (see also subsequent work in other holographic models \cite{Goykhman:2012vy,Brattan:2012nb,Jokela:2012se,Jokela:2015aha,Itsios:2015kja,Itsios:2016ffv}). We have indeed numerically verified this expectation in the current system. In Fig.~\ref{zero_sound_gap_B}, we have restricted to low-temperature regime and plotted numerical data against the gap in (\ref{eq:omatBne0}). We nicely see that the analytic results conform with full numerical analysis for sufficiently small magnetic field strengths. We have also presented the dispersions in Fig.~\ref{zero_sound_gap_B} following (\ref{eq:omatBne0withk}), which also match the numerics.

Like before, we can obtain the next-to-leading order corrections, where we encounter an imaginary part of $\omega$. A straightforward computation yields
\beq
{\rm{Im}}\, \delta\omega = -\frac{\pi(c_s^2k^2+\omega_0^2)^{\frac{\xi}{z}-1}(c_s^2k^2+2 \omega_0^2)}{2 d I_0 (2z)^{\frac{2\xi}{z}}\Gamma\left(\frac{\xi}{z}+\frac{1}{2}\right)^2} \ .
\eeq

\subsection{Crossover from hydrodynamic to collisionless regime}

We found the zero sound mode analytically in the $T=0$ case and the diffusion mode in the $T\neq 0$ case. It is natural to expect that for some range of temperatures, both modes would coexist although dominate at different momenta. The regime close to the $T=0$ is called the collisionless (quantum) regime, where the dominant physics are captured by the collective sound mode-like excitation appearing as the pole in the density-density correlator. In usual Landau-Fermi liquids, such a mode is visible as due to oscillations of the Fermi surface of the underlying interacting fermions. The system is pretty robust to temperature variations. For example, while the attenuation of the zero sound gets corrections from thermal effects, the speed of zero sound is quite insensitive to these. At sufficiently high temperatures, there is a phase transition to a hydrodynamic regime. The scaling of this phase transition point strongly depends on parameters of the system. 

\begin{figure}[ht]
\center
 \includegraphics[width=0.40\textwidth]{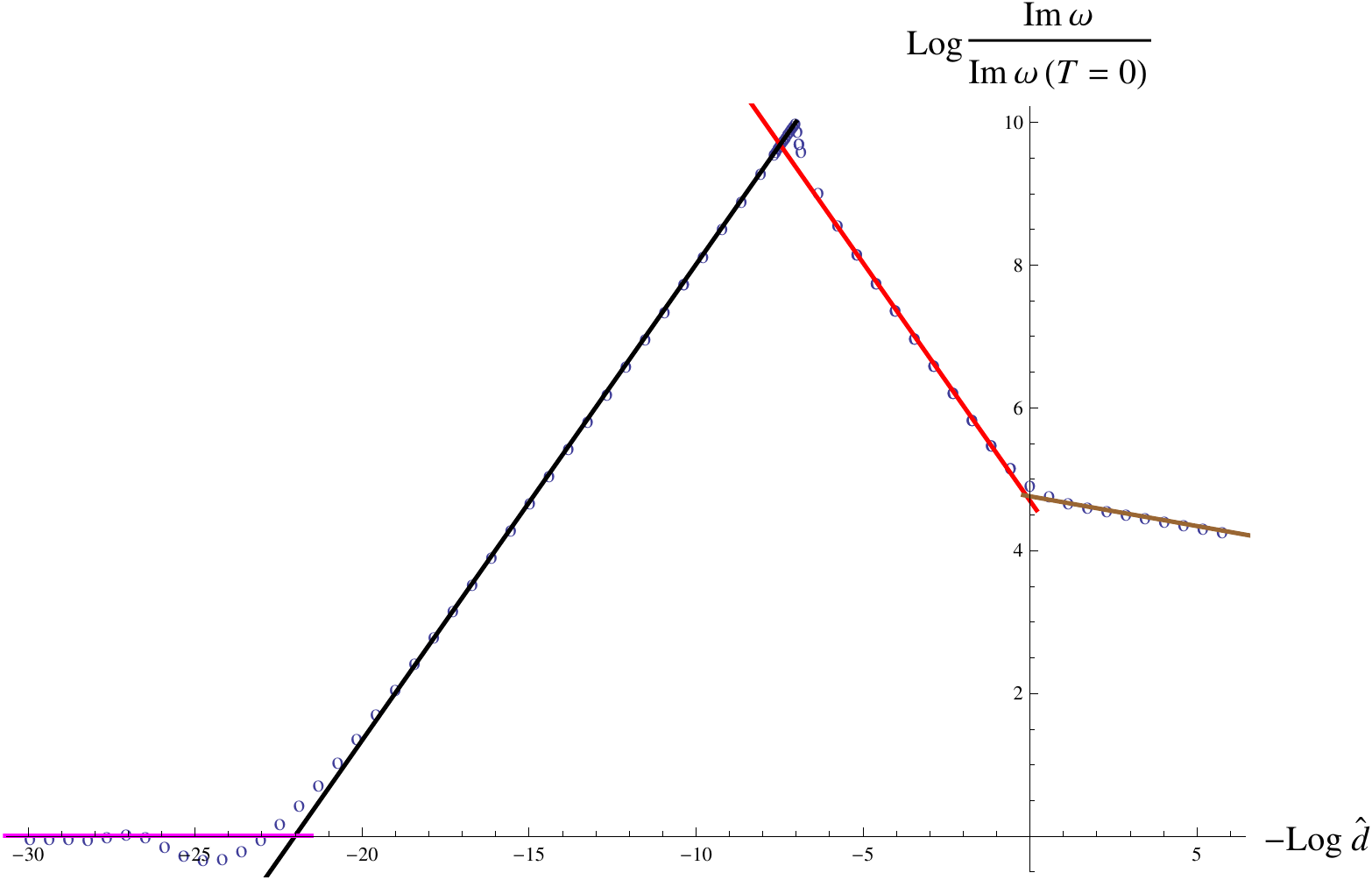}
   \caption{The damping rate of the most dominant mode is plotted against increasing temperature in the logarithmic scale with fixed $\frac{\hat k}{\hat d^{1/(q\xi)}} = 0.01$ and at $B=0$ for $q=3,\ z=1.5$, and $\xi=2$. We have chosen to normalize the imaginary part of the frequency at $T=0$. The collisionless $T\sim 0$ regime corresponds to the left-most part of the plot, where the zero sound is the dominant mode. The zero slope means that its imaginary part is pretty robust to temperature variations up until the system enters in the thermal collisionless regime. The transition to the hydrodynamic regime is marked as the highest point of the plot, after of which the diffusion mode takes over. The slopes of the lines are $0,2/3,-2/3,1/12$. We find a very accurate match with the numerics and the analytic predictions.}
\label{damping_rate}
\end{figure}

In Fig.~\ref{damping_rate} we have plotted the imaginary value of the most dominant mode as a function of increasing temperature. In this plot, we notice the different regimes. To the far-left is the collisionless quantum regime, which transitions to thermal collisionless regime, where the attenuation of the zero sound assumes (positive) exponential scaling with the temperature. At sufficiently large temperature, we find a sharp transition to the hydrodynamic regime, where the physics is dominated by the diffusion mode, whose decay rate scales with negative power of the temperature.

\begin{figure}[ht]
\center
 \includegraphics[width=0.40\textwidth]{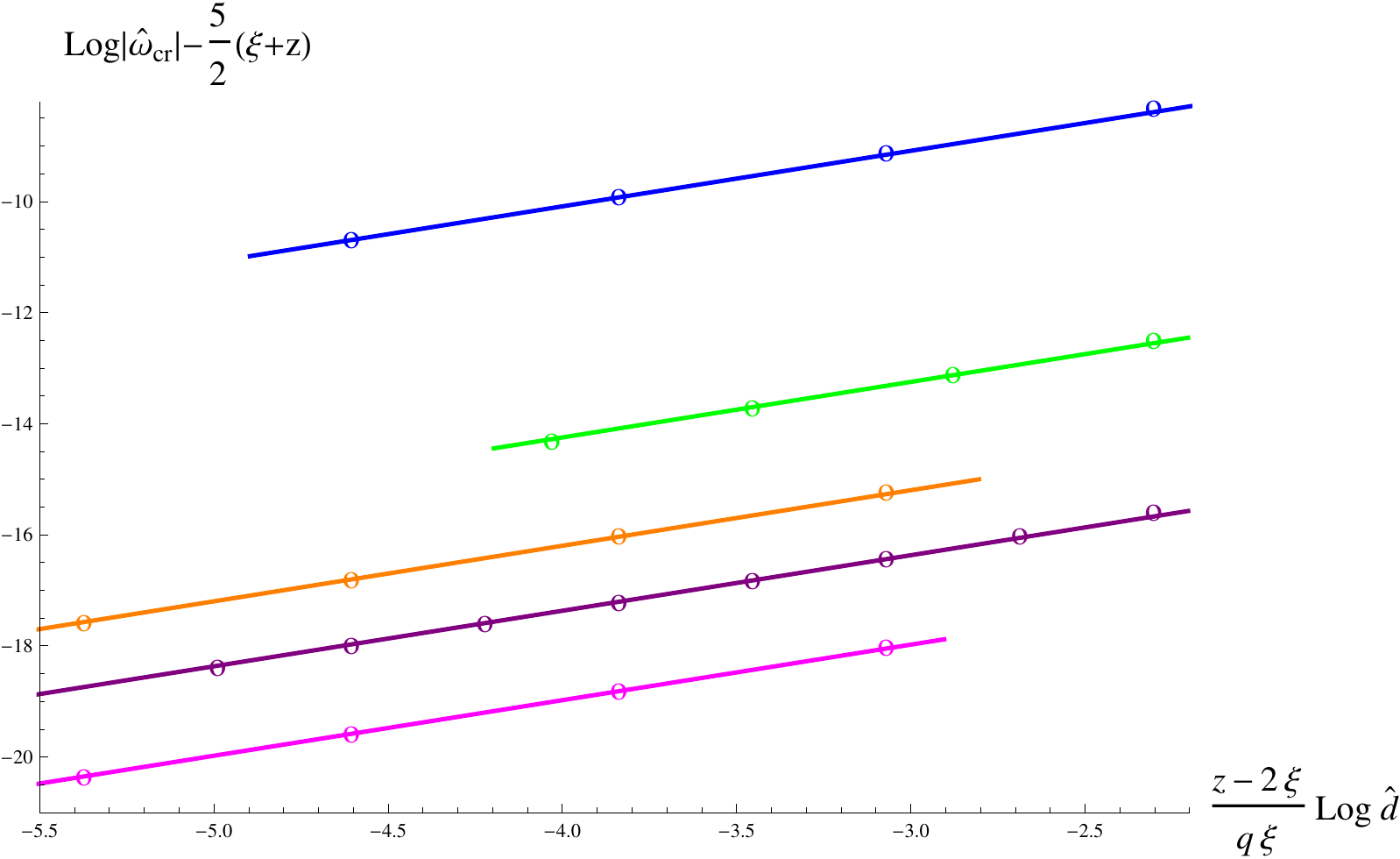}
 \qquad\qquad
  \includegraphics[width=0.40\textwidth]{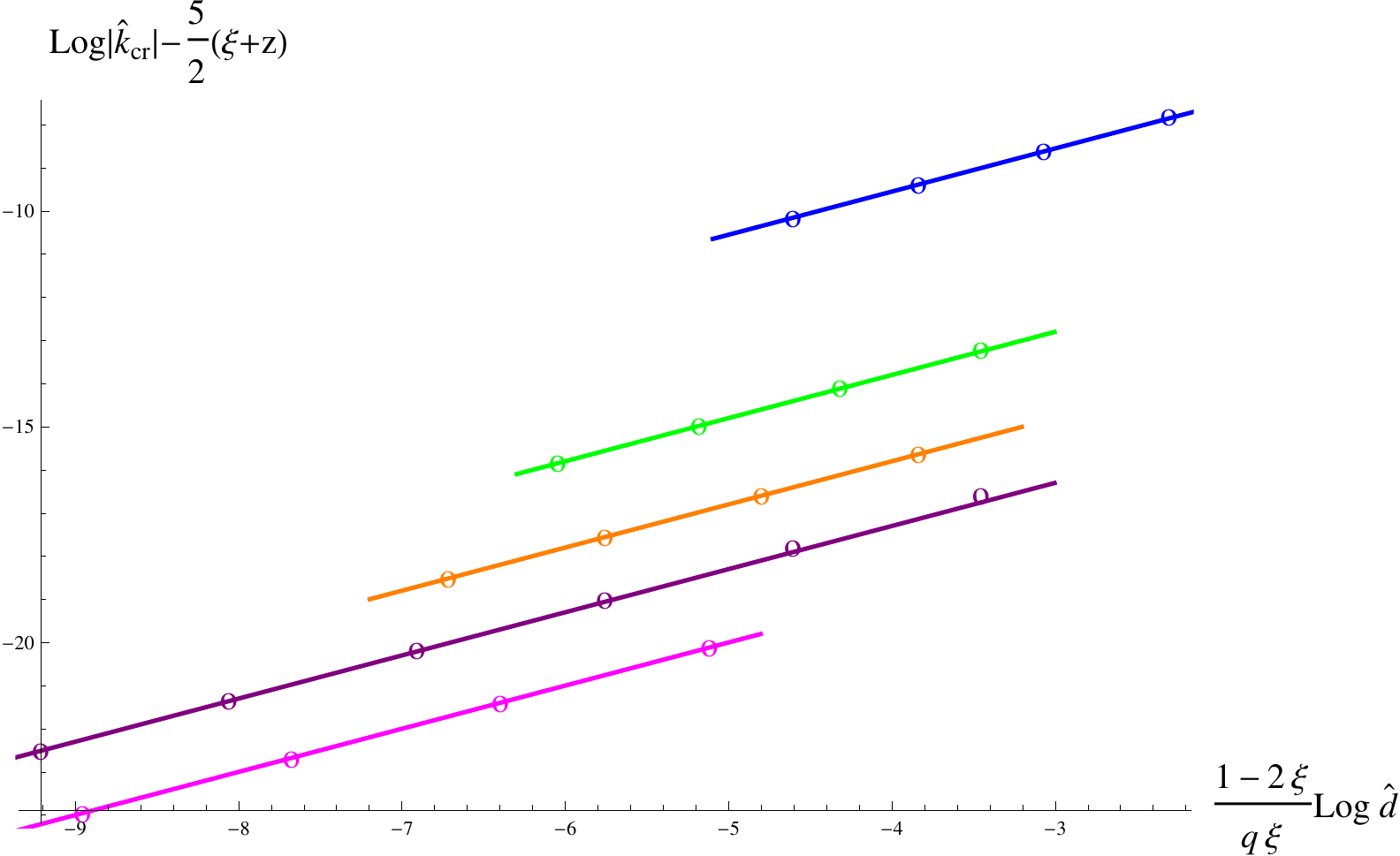}
\caption{We depict the frequencies $|\hat \om_{cr}|$ (left) and wave vectors $\hat k_{cr}$ (right) corresponding to the transition point from the hydrodynamical to the collisionless regime for various values of the parameters against increasing temperature (from left to right) in a logarithmic scale. We have chosen to plot the various lines that we fit to numerical data (points) such that all the cases have slopes corresponding to 1. Notice, that we have included a constant vertical shift to separate the lines. The lines correspond to $(q=3,\xi=1,z=1)$, $(q=4,\xi=2,z=2)$, $(q=4,\xi=3,z=2)$, $(q=3,\xi=2,z=3)$, and $(q=3,\xi=3,z=3)$ (top-down). We conclude that the scaling laws (\ref{omega_k_cr}) are faithfully captured by the numerical data.}
\label{omega_k_crit}
\end{figure}

The aim of this subsection is to extract the scaling law of the phase transition point from numerical results at $B=0$. We were successful in predicting the quite complicated form of the scaling exponents (see equation (\ref{omega_k_cr}) below) and they match the numerical results very accurately. More explicitly, we work at finite temperature and small momenta, and set up the numerics to finding a transition point where a purely imaginary diffusive mode transforms into a zero sound mode. A typical representation of the dispersions is as in Fig.~\ref{Dispersion_relation}, where the transition point from the hydrodynamic diffusive mode merging with another purely imaginary mode to form a pair of complex sound modes is clearly visible.  At this point the angular frequency and momentum take what we call critical values: ($\hat k_{cr}$, $\hat \omega_{cr}$). The critical values depend on the temperature and chemical potential of the system. Using numerical analysis, we determined how these parameters depend on the temperature and on the chemical potential:
\beq
\omega_{cr} \sim \frac{T^{\frac{2 \xi}{z}}}{\mu_0^{\frac{2\xi-z}{z+2\xi-2}}}\,\,,
\qquad\qquad
k_{cr} \sim \frac{T^{\frac{2 \xi}{z}}}{\mu_0^{\frac{2\xi-1}{z+2\xi-2}}} \ .
\label{omega_k_cr}
\eeq
Here, we have used the zero temperature chemical potential, $\mu_0$. 
We emphasize that the scaling relations in (\ref{omega_k_cr}) are only expected to hold at sufficiently low temperatures.
When we set both $z$ and $\xi$ to unity, the scaling becomes $\frac{T^2}{\mu_0}$, which is the expected result for conformal background generated by D3-branes, see {\emph{e.g.}}, \cite{Jokela:2015aha}. In Fig.~\ref{omega_k_crit} we plot the transition values as functions of increasing temperature. The predicted scaling laws (\ref{omega_k_cr}) match the numerics extremely well.

\section{Alternative quantization}
\label{sec:alternative}

In this section, we consider the special case $q=2$ as it allows us to move away from the traditional Dirichlet boundary condition and consider a mixed Dirichlet-Neumann boundary conditions. 
We will briefly review the necessary notation and the background of this following \cite{Jokela:2013hta} (see also \cite{Jokela:2014wsa,Brattan:2013wya,Brattan:2014moa,Jokela:2015aha,Itsios:2015kja,Itsios:2016ffv}), which contains a more detailed analysis of the alternative quantization. For a complementary discussion in implementing alternative boundary conditions in holography, see \cite{Ihl:2016sop}. The approach in \cite{Ihl:2016sop} is slightly more abstract, but has the benefit of straightforward generalization off-shell as well as an unambiguous way of computing the thermodynamic potentials of the anyonized system. 

Consider our original action. When we vary the gauge fields and impose the equations of motion, we have the boundary term
\beq
\delta S_D = \int_{\rm{bdry}} J^{\mu}\delta A_{\mu} \ ,
\eeq
where $J^{\mu}$ is interpreted as the conserved current of the boundary theory. In the previous sections, we have demanded that this be zero, effectively requiring $\delta A_{\mu}=0$ at the boundary, \ie, the Dirichlet boundary condition. The modified boundary conditions are achieved by adding additional boundary terms to the action.

As $\partial_{\mu}J^{\mu}=0$, we can write $J_{\mu}= \frac{1}{2\pi}\varepsilon_{\mu\nu\lambda}\partial^{\nu}v^{\lambda}$, where $v^\mu$ is an arbitrary vector corresponding to the gauge redundancy.\footnote{For discussion on how to fix this, see \cite{Ihl:2016sop}. The freedom parameterized by $v^\mu$, does not enter in the pole structure, but will be important if one wishes to compute thermodynamic potentials.} In the following, we also write $B_{\mu} = \frac{1}{2\pi}\varepsilon_{\mu\nu\lambda}\partial^{\nu}A^{\lambda}$. Now, the most general action we can write while still retaining the original equations of motion is
\beq\label{eq:alternativebc}
S = S_D + \frac{1}{2\pi}\int_{bdry} \left[a_1\varepsilon_{\mu\nu\lambda}A^{\mu}\partial^{\nu}v^{\lambda} +a_2\varepsilon_{\mu\nu\lambda}A^{\mu}\partial^{\nu}A^{\lambda} + a_3\varepsilon_{\mu\nu\lambda}v^{\mu}\partial^{\nu}v^{\lambda}\right] \ .
\eeq
Variation of both $v^{\mu}$ and $A^{\mu}$ gives us
\beq
\delta S = \int_{\rm{bdry}} (a_s J_{\mu}+b_s B_{\mu})(c_s\delta v^{\mu}+d_s\delta A^{\mu}) \ ,
\eeq
where
\beq
a_sd_s = 1+a_1\ ,\quad b_sc_s = a_1\ ,\quad b_sd_s=2a_2\ ,\quad a_sc_s = 2 a_3\ .
\eeq
Notice that  $a_sd_s-b_sc_s=1$, as the transformations can be identified with the elements in $SL(2,\mathbb{R})$. From the variation of $S$, we can see that our boundary condition has become
\beq
c_s\delta v^{\mu}+d_s \delta A^{\mu}\Big |_{r\to \infty}=0 \qquad\Leftrightarrow\qquad c_s\delta J^{\mu}+d_s \delta B^{\mu}=0 \ .
\eeq
In addition, we can read the new current corresponding to the new boundary conditions,
\beq
J_\mu^* =a_s J_{\mu}+b_s B_{\mu} \ .
\eeq

Having reviewed some basic renditions from adding boundary terms in the action, we will now focus on the pole structure of the Green's functions. The effect of using mixed boundary conditions for the electromagnetic fields can be summarized in the following condition
\beq
\lim\limits_{r\to \infty}\left[\mathfrak{n}r^{\lambda}f_{r\mu}-\frac{1}{2}\varepsilon_{\mu\alpha\beta}f^{\alpha\beta}\right]=0 \ , \quad \mu=t\,,x\,,y \ .
\eeq
The indices in the second term have been raised with the Minkowski metric $\eta_{\mu\nu}$ whose temporal component has been scaled with $r^{2(z-1)}$ to take into account the Lifshitz scaling. The parameter $\lambda$ is determined from the scaling properties of the classical fields. The parameter $\mathfrak{n}$ measures the state of mixedness of the boundary conditions in comparison to the Dirichlet boundary condition. When $\mathfrak{n}=0$, we recover the Dirichlet boundary condition and with $\mathfrak{n}\to\infty$, we asymptotically approach the Neumann boundary condition. Equivalently, we could use
\beq
\lim\limits_{r\to \infty}\left[r^{\lambda}f_{r\mu}-\frac{\mathfrak{m}}{2}\varepsilon_{\mu\alpha\beta}f^{\alpha\beta}\right]=0 \ ,
\eeq
where $\mathfrak{m}=1/\mathfrak{n}$.

More explicitly, these conditions are for $\mu=t,\,x$:
\beq
\lim\limits_{r\to \infty}\left[\mathfrak{n}r^{\lambda}a'_{t}-ik a_y\right]=0,\quad \lim\limits_{r\to \infty}\left[\mathfrak{n}r^{\lambda}a'_{x}+i\omega r^{2(1-z)}a_y\right]=0 \ . \label{mixed_at_ax}
\eeq
We can use the relation \eqref{at_ax_E}, to write the two conditions \eqref{mixed_at_ax} in a single condition in terms of $a_y$ and $E$:
\beq
\lim\limits_{{r\to \infty}}\left[{\mathfrak{n}r^{\lambda+2(z-1)}E'\over \omega^2 -k^2r^{2(z-1)}}+ia_y\right]=0 \ .
\eeq
Finally, for $\mu=y$:
\beq
\lim\limits_{r\to \infty}\left[\mathfrak{n}r^{\lambda}a'_{y}-ir^{2(1-z)}E\right]=0 \ .
\eeq

\subsection{Zero sound}

We now apply the boundary conditions to our solution for the equations of motion for the zero sound mode with magnetic field turned on, \eqref{E-sol-zs-small-k}, \eqref{ay-sol-zs-small-k}. Setting $\lambda=3-z$, the boundary conditions become
\beq
\left\{\begin{array}{l}\mathfrak{n} c_E+ i a_y^{(0)}=0\\ \mathfrak{n}c_y-iE^{(0)} =0\end{array}\right. \ .
\eeq
Implementing these conditions to match the two expansions done in different orders, the matrix relation \eqref{E0_ay0_B} becomes,
\beq
\begin{pmatrix}
E^{(0)}+i\mathfrak{n}c_y\\ \\ a_y^{(0)}-i\mathfrak{n}c_E
\end{pmatrix}
\,=\,
\begin{pmatrix}
\omega^2\,I_0\,-\,k^2\,J_0\,-\,{\omega^{1+{2\xi\over z}}\over (z-2\xi)\,c\,d}
 &&&-i\left({B\over d}-\mathfrak{n}\right)\\
  {} & {} \\
 i\left({B\over d}-\mathfrak{n}\right) &&& 
 I_0-{\omega^{1+{2\xi\over z}}\over (z-2\xi)\,c\,d}
  \end{pmatrix}\,
 \begin{pmatrix}
c_E\\ \\c_y
\end{pmatrix} \ .
\eeq

\begin{figure}[ht]
\center
 \includegraphics[width=0.50\textwidth]{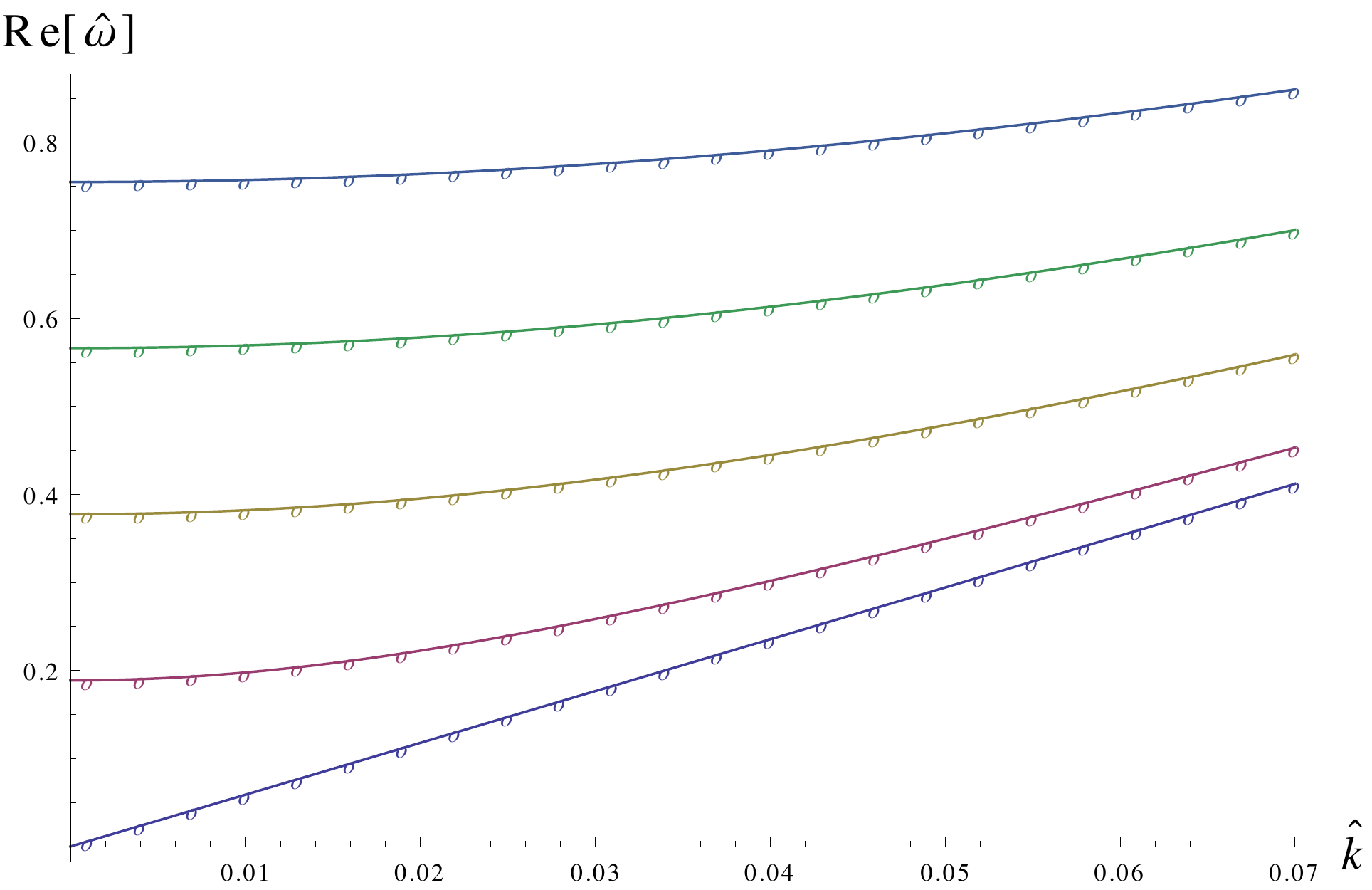}
   \caption{A comparison of real part of the zero sound mode with a non-zero $B$ and alternative quantization. The parameters are $\hat d=10^4$, $\hat B=1000$, $q=2$, $z=1.5$, and $\xi =2$. We vary the alternative quantization parameter $\mathfrak{n}=0, 250/10^4, 500/10^4, 750/10^4, 1000/10^4$ (top-down).}
\label{zero_sound_alternative}
\end{figure}

For the non-trivial solution, we require that the determinant of the matrix vanishes. We see that the dispersion relation can be obtained from our previous solution by a simple shift $B\to B - d\mathfrak{n}$. With this, we can have a gapless dispersion relation even with the magnetic field. More explicitly, the dispersion relation in leading order is
\beq\label{eq:alterzero}
\omega = \pm\sqrt{c_s^2\,k^2+\left(\frac{4\xi (B-d\mathfrak{n}) d^{\frac{z-2\xi}{2\xi}} }{B\left(\frac{2\xi-z}{4\xi},\frac{z}{4\xi}\right)}\right)^2} \ .
\eeq
The next-to-leading order contribution can also be found by modifying the previous results accordingly.
The effect of mixing the boundary conditions agrees exactly with \cite{Jokela:2015aha}.
We also compare the analytic expression in (\ref{eq:alterzero}) with the numerics in Fig.~\ref{zero_sound_alternative}. We have chosen to present the dispersion of (\ref{eq:alterzero}) for a given set of parameters, but the numerical match is very good for other choices $\xi$ and $z$, as well.

\subsection{Diffusion constant}

For the diffusion constant, we consider the case without a magnetic field.\footnote{The calculation could be generalized to finite, and small, $B$. To streamline the discussion, we have decided not to include it in as it would be a very long illustration.} For this calculation, we need the solution to the finite temperature equations of motion for the transverse field $a_y$. The calculation can be found in Appendix \ref{appendix:transverse}. Applying these solutions to the new boundary conditions and setting $\lambda =3-z$ we get
\beq
\left\{\begin{array}{l} \frac{\mathfrak{n}}{k^2}c_E + i a_y^{(0)}=0\\
\mathfrak{n}\alpha_0 Z(\omega,k)-iE^{(0)} =0 \end{array} \right. \ ,
\eeq
where $\alpha_0$ and $Z(\omega,k)$ appear in (\ref{eq:defofZ}). We can reduce this to a single equation,
\beq
E^{(0)}=\frac{\mathfrak{n}^2 Z(\omega,k)}{k^2} c_E 
\eeq 
which we can plug into the matching condition for $E$ \eqref{matching_linear_diff}, to solve for the dispersion relation
\be
\omega = -iD_{\mathfrak{n}}k^2\ ,\qquad D_{\mathfrak{n}}=D^*+\frac{(D-D^*)r_H^{4\xi}}{r_H^{4\xi}+\mathfrak{n}^2(d^2+r_H^{4\xi })} \ ,
\ee
where we have employed $D^*$ from \eqref{dstar}. We see that the diffusion constant starts from $D$ and ends up at $D^*$ as we vary $\mathfrak{n}^2$ from $0$ to $\infty$. This agrees qualitatively with the results in \cite{Jokela:2015aha,Brattan:2013wya}. In these papers, the authors obtained the results using the properties of $SL(2,\mathbb{R})$ transformations.

\subsection{Conductivities}

We now make use of our previous results to determine the conductivities of the system. We will stick to the case with no magnetic field. In general, the conductivity can be calculated as follows
\beq
\sigma_{ij}(\omega) = \frac{1}{i\omega}\langle J_i(-\omega,0)J_j(\omega,0)\rangle \ .
\eeq 
As we wish to compute the whole Green's function and not focus solely on the poles, we need to analyze the on-shell action. We first focus on the original Lagrangian written in (\ref{eq:fluctuationLagrangian}). We will then proceed to adding appropriate boundary terms, as discussed in the beginning of this section, to extracting the Green's function for the anyonized system. To simplify the procedure, we will restrict to $SL(2,\mathbb{Z})$ transformations. For complementary discussion on computing the alternatively quantized conductivities, see \cite{Brattan:2013wya,Itsios:2016ffv,Jokela:2015aha,Ihl:2016sop}.

\subsubsection{Conductivities with the Dirichlet boundary condition}

First, we compute the conductivities at the low frequency limit using the Dirichlet boundary conditions. Notice that $q$ can take any value in this subsection, whereas in the following subsection, where we consider the alternative quantization, it is assumed to be 2. The part of the Lagrangian density involving $E$ is
\beq
\mathcal{L}\propto -\frac{\mathcal{N}}{2}G^{tt}\frac{\sqrt{g_{rr}|g_{tt}|}}{\sqrt{H+d^2}}H\left[-G^{rr}v^2a_x'^2+G^{rr}a_t'^2-G^{xx}E^2 \right]
\eeq
which can be easily partially integrated to give, after using the constraint (\ref{at_ax_E}),
\beq
S_{\rm on-shell}=-\frac{\mathcal{N}}{2}\int {\rm d}\omega\,{\rm d}^{q}k\,  G^{tt}G^{rr}\frac{\sqrt{g_{rr}|g_{tt}|}}{\sqrt{H+d^2}}H\frac{E(-\omega,-k) E'(\omega,k)}{\omega^2-k^2u^2}\Big|_{r\to\infty}.
\eeq

Using the solutions to the equations of motion at zero temperature, we have
\beq
S_{\rm{on-shell}}=\frac{\mathcal{N}}{2}\int {\rm d}\omega\, {\rm d}^{q}k\, E^{(0)}(-\omega,-k)c_E(\omega,k).
\eeq
Using the matching condition \eqref{E0_ay0_B}, we get
\bear
\langle J_x(-\omega,-k)J_x(\omega,k)\rangle \!&=&\! \frac{\delta^2 S_{\rm{on-shell}}}{\delta a_x(\omega,k)\delta a_x(-\omega,-k)}\\
      &= &\frac{{\rm d} E(\omega,k)}{{\rm d} a_x(\omega,k)}\,\frac{{\rm d} E(-\omega,-k)}{{\rm d} a_x(-\omega,-k)}\,\frac{\delta^2S_{\rm{on-shell}}}{\delta E(\omega,k)\delta E(-\omega,-k)}\\ 
&=& -\frac{\omega^2 \mathcal{N}}{\Big(\omega^2\,I_0-k^2\,J_0\,-\,{\omega^{1+{2\xi\over z}}\over (z-2\xi)\,c\,d}\,\Big)} \ .
\eear
At the low-frequency limit, the conductivity is
\beq
\sigma_{xx}(\omega) = i\mathcal{N}\begin{cases}\frac{1}{I_0}\omega^{-1} & {\rm{if}}\,z<2\xi\\
  -(z-2\xi)cd\omega^{-\frac{2\xi}{z}} & {\rm{if}}\, z>2\xi\end{cases} \ , \ \ T = 0 \ .
\eeq
In addition, when $z=2\xi$, there is also a logarithmic contribution.

When we consider the system at finite temperature, we get
\beq
S_{\rm{on-shell}}=\frac{\mathcal{N}}{2}\int {\rm d}\omega {\rm d}^{q}k\, \frac{E^{(0)}(-\omega,-k)c_E(\omega,k)}{k^2}.
\eeq
and using the matching condition \eqref{matching_linear_diff}, we have the two-point function
\beq
\langle J_x(-\omega,-k)J_x(\omega,k)\rangle = \mathcal{N}\frac{\omega^2 r_H^{\xi(q-2)}\sqrt{1+d^2r_H^{-2\xi q} } }{k^2 D-i \omega} \ ,
\eeq
from which we get the DC conductivity (for any $q$),
\beq
\sigma_{xx}(\omega) = \frac{1}{i \omega}\langle J_x(-\omega,0)J_x(\omega,0)\rangle = \mathcal{N} r_H^{\xi(q-2)}\sqrt{1+d^2r_H^{-2\xi q} } \ .\label{dc-conductivity}
\eeq

Due to the absence of the magnetic field, there is no coupling between $E$ and $a_y$, and therefore the conductivity tensor is diagonal, $\sigma_{ij}\propto\delta_{ij}$. 
However, in the following we will consider the alternative boundary conditions jazzing up the situation. These boundary conditions have the effect that, while the equations of motion remain intact, they  link $E$ and $a_y$ together, and will therefore generate off-diagonal elements in the conductivity tensor.

\subsubsection{Effect of alternative quantization on conductivity}

In the following, we will calculate the effect of the $S$ and $T$ transformations of $SL(2,\mathbb{Z})$ on the conductivity tensor and also a more general $SL(2,\mathbb{Z})$ transformation. We will only consider the case $k=0$.

Let $a(\omega)=(a_t(\omega),a_x(\omega),a_y(\omega))^T$ be a vector of the gauge fields on the boundary. Also, let $M^{\mu\nu}(\omega)$ be a $3\times 3$ matrix with the property $M^{\mu\nu}(-\omega)=M^{\nu\mu}(\omega)$. Now, for the Dirichlet boundary condition, let the action on the boundary be
\beq\label{eq:action-dirichlet}
S^{(2)}_{\rm{on-shell}}=\frac{1}{2}\int {\rm d}\omega \, a_{\mu}(-\omega) M^{\mu\nu}(\omega) a_{\nu}(\omega)\ .
\eeq
The current terms with the Dirichlet boundary conditions are
\beq
J^{\mu}(\omega)=\frac{\delta S^{(2)}_{\rm{on-shell}}}{\delta a_{\mu}(-\omega)}=M^{\mu\nu}(\omega) a_{\nu}(\omega)\ ,
\eeq
or expressing them in terms of the vector $v^\mu$ defined earlier around (\ref{eq:alternativebc}), we have
\beq
J^t(\omega)=0 \ , \quad J^x(\omega) = -\frac{i\omega}{2\pi}v^y(\omega)\ ,\quad J^y(\omega) =\frac{i\omega}{2\pi}v^x(\omega)  \ .
\eeq
Restricting ourselves to the spatial components of $M$, we can express the relation above of the spatial components neatly with a matrix $S$:
\bear
\vec{J}(\omega) &=& -\frac{i\omega}{2\pi}\left(\begin{array}{cc} 0 & 1\\ -1 & 0\end{array}\right)\vec{v}(\omega)=-\frac{i\omega}{2\pi}S\cdot \vec{v}(\omega)\\
\vec{v}(\omega)&=&-\frac{i 2\pi}{\omega} S\cdot \vec{J}=-i\frac{2\pi}{\omega}S\cdot M(\omega)\cdot \vec{a} \ .
\eear
The new boundary conditions can be written in a matrix form
\bear
c_s \vec{v}(\omega)+d_s \vec{a}(\omega)&=& \left(-\frac{i c_s2\pi}{\omega}S\cdot M+d_s\right)\vec{a}(\omega)\equiv \vec{H}(\omega)=0 	\\
\vec{a}(\omega)&=&\left(-\frac{i c_s2\pi}{\omega}S\cdot M(\omega)+d_s\right)^{-1}\vec{H}(\omega) \ .
\eear

Consider the additional boundary terms given at the beginning of this section around (\ref{eq:alternativebc}). First, we consider a $T^K$ transformation, for which $a_s=d_s=1$, $b_s=K$ and $c_s=0$. Thus, the new additional boundary terms will be (with $\vec{a}=\vec{H}$)
\beq
-\frac{K}{4\pi}\int {\rm d}\omega\,i\omega\vec{a}^T(-\omega)\cdot S \cdot \vec{a}(\omega)\ ,
\eeq
thus the full action integral is modified to the following form
\beq
S_{T^K}=\frac{1}{2}\int \vec{a}^T(-\omega)\left(M-i\frac{K\omega}{2\pi}S\right)\vec{a}(\omega)\ ,
\eeq
which gives us the transformed conductivities
\bear
\sigma^*_{L,T^K}(\omega)&=&\frac{M_{xx}(\omega)}{i\omega}=\sigma_L(\omega) \\
\sigma^*_{H,T^K}(\omega)&=&\frac{M_{xy}}{i\omega}-\frac{K}{2\pi}=\sigma_H(\omega)-\frac{K}{2\pi} \ .
\eear

The $S$ transformation can be done similarly although the calculation is slightly more involved. This time $a_s=d_s=0$ and $b_s=1=-c_s$. The additional boundary term is
\bear
&&-\frac{i}{2\pi}\int {\rm d}\omega\, \omega\vec{a}^T(-\omega)\cdot S \cdot \vec{v}(\omega)=-\int {\rm d}\omega \, \vec{a}^T(-\omega)\cdot  M(\omega) \cdot \vec{a}(\omega)\\
&=&-\int {\rm d}\omega\, \vec{H}^T(-\omega)S^T M^{-1}(\omega) S \vec{H}(\omega)\ .
\eear
Thus, the full action takes the form
\bear
-\frac{1}{2}\int {\rm d}\omega\,\frac{\omega^2}{4\pi^2}  \vec{H}^T(-\omega) S^T M^{-1}(\omega) S \vec{ H}(\omega)\ ,
\eear
from which we can compute the conductivities
\beq
\sigma^*_{L,S}=\frac{1}{(2\pi)^2}\frac{\sigma_L}{\sigma_L^2+\sigma_H^2}\quad , \quad \sigma^*_{H,S}=\frac{1}{(2\pi)^2}\frac{-\sigma_H}{\sigma_L^2+\sigma_H^2} \ .
\eeq

The above results agree with previous results in the literature. Let us finally note, that the general transformation for the $ST^K$ reads
\beq
\sigma^*_{L,ST^K}=\frac{1}{(2\pi)^2}\frac{\sigma_L}{\sigma_L^2+\frac{K^2}{4\pi^2}}\quad , \quad \sigma^*_{H,ST^K}=\frac{1}{(2\pi)^3}\frac{K}{\sigma_L^2+\frac{K^2}{4\pi^2}} \ . \label{eq:conductivity-STK}
\eeq

Recall the relation $\mathfrak{n}^{-1}=\mathfrak{m}=\frac{K}{2\pi \mathcal{N}}$. It is noteworthy that we recover the longitudinal conductivity of Dirichlet quantization by multiplying the expression of $\sigma^*_{L,ST^K}$ with $\frac{4\pi^2}{\mathfrak{n}^2}$ and taking the limit $\mathfrak{n}\to 0$. This is due to the fact that the alternative quantization not only alters position of poles in the Green function but also the source fields and currents. For more discussion on this limiting procedure, we refer the reader to \cite{Brattan:2013wya}.

\subsection{Einstein relation}
Let us illustrate that the Einstein relation holds in our system. The Einstein relation states that the diffusion coefficient, charge susceptibility, and longitudinal DC conductivity are related by
\beq
D \chi = \sigma_L \ ,
\eeq
where $\chi=\left(\frac{\partial\rho}{\partial\mu}\right)_T$. Recalling our previous results in the absence of magnetic field, (\ref{eq:chem}), (\ref{hatD_zeroB}), and (\ref{dc-conductivity}), we can indeed verify that the Einstein relation holds exactly for all $z$, $\xi$, and $q$ when the system obeys Dirichlet boundary conditions. We revisit this claim in the presence of $B$ in section \ref{sec:comments}.

When we consider the case of alternative quantization, we face the obstacle of defining what we mean with the chemical potential and charge density. However, we can easily circumvent most of this by considering the alternatively quantized current-current correlator \cite{Brattan:2014moa},
\beq
\chi^* = \langle J_t^*(-k) J_t^*(k)\rangle\bigg|_{\omega=0,{\vec k} \ll 1} \ .
\eeq
The computation of this quantity is similar to the above computation of conductivity. 

We start working with the action with the Dirichlet boundary condition
\beq
S^{(2)}_{\rm{on-shell}}=\frac{1}{2}\int {\rm d}k \, a_{\mu}(-k) M^{\mu\nu}(k) a_{\nu}(k)\,
\eeq
which bears a striking resemblance to \eqref{eq:action-dirichlet} with the difference that we have set $\omega=0$ and we are working with $k\equiv k_x$ dependence only. Following the steps we took previously, we can compute the currents and relate the function $v$ to the currents
\beq
\left(\begin{array}{c} v_t\\ v_y \end{array}\right)= -\frac{2\pi i}{k}S\cdot \left(\begin{array}{c} -J_t\\ J_y \end{array}\right)= -\frac{2\pi i}{k}S\cdot M\cdot\left(\begin{array}{c} a_t\\ a_y \end{array}\right) \ ,
\eeq
with which we can write the modified boundary conditions,
\bear
\left(-\frac{c_s 2\pi i}{k}S\cdot M + d_s\right)\left(\begin{array}{c} a_t\\ a_y \end{array}\right)\equiv \left(\begin{array}{c} H_t\\ H_y \end{array}\right)  = 0 \\
\left(\begin{array}{c} a_t\\ a_y \end{array}\right) = \left(-\frac{c_s 2\pi i}{k}S\cdot M + d_s\right)^{-1}\left(\begin{array}{c} H_t\\ H_y \end{array}\right) \ .
\eear

We consider the $T^K$ transformation first. This transformation does not affect the susceptibility as it only modifies the non-diagonal terms:
\beq
M_{ty}\to M_{ty}-i\frac{kK}{2\pi} \ .
\eeq 
We can also recall from our previous discussion that the $T$ transformation does not affect the longitudinal conductivity or the diffusion coefficient either, trivially satisfying the Einstein relation.

Then, consider the effect of $S$ transformation. Following the steps taken with conductivity, we can easily see that
\beq
\chi^*_S = -\frac{k^2}{4\pi^2}\frac{M_{tt}}{M_{tt}M_{yy}+M_{ty}^2} \ .
\eeq
Thus, after an $ST^K$ transformation, the susceptibility of our system is
\beq
\chi^*_{ST^K}=\frac{-1}{4\pi^2}\frac{\chi}{-\chi D^*\sqrt{1+d^2r_H^{-2\xi q}} -\frac{K^2}{4\pi^2}}\label{eq:susceptibility-STK} \ ,
\eeq
where we used the coefficient $D^*$ from \eqref{dstar} and $\chi$ is the original susceptibility. Combining all of our results from discussions of conductivity and diffusion with alternative quantization, we can verify that the Einstein relation is satisfied in this case, too.


\section{Comments: DC conductivity with a magnetic field}\label{sec:comments}

Encouraged by the successful confirmation of the Einstein relation above, we attempt to obtain an expression for longitudinal DC conductivity by considering the Einstein relation true even in the presence of $B$. 
Both the numerical and analytical evidence for the Einstein relation was overwhelming in the absence of the magnetic field in the preceding sections, so we now take the next logical step and predict the DC conductivity of the (normally quantized) system also in the presence of the magnetic field $B$.

\begin{figure}[ht]
\center
 \includegraphics[width=0.40\textwidth]{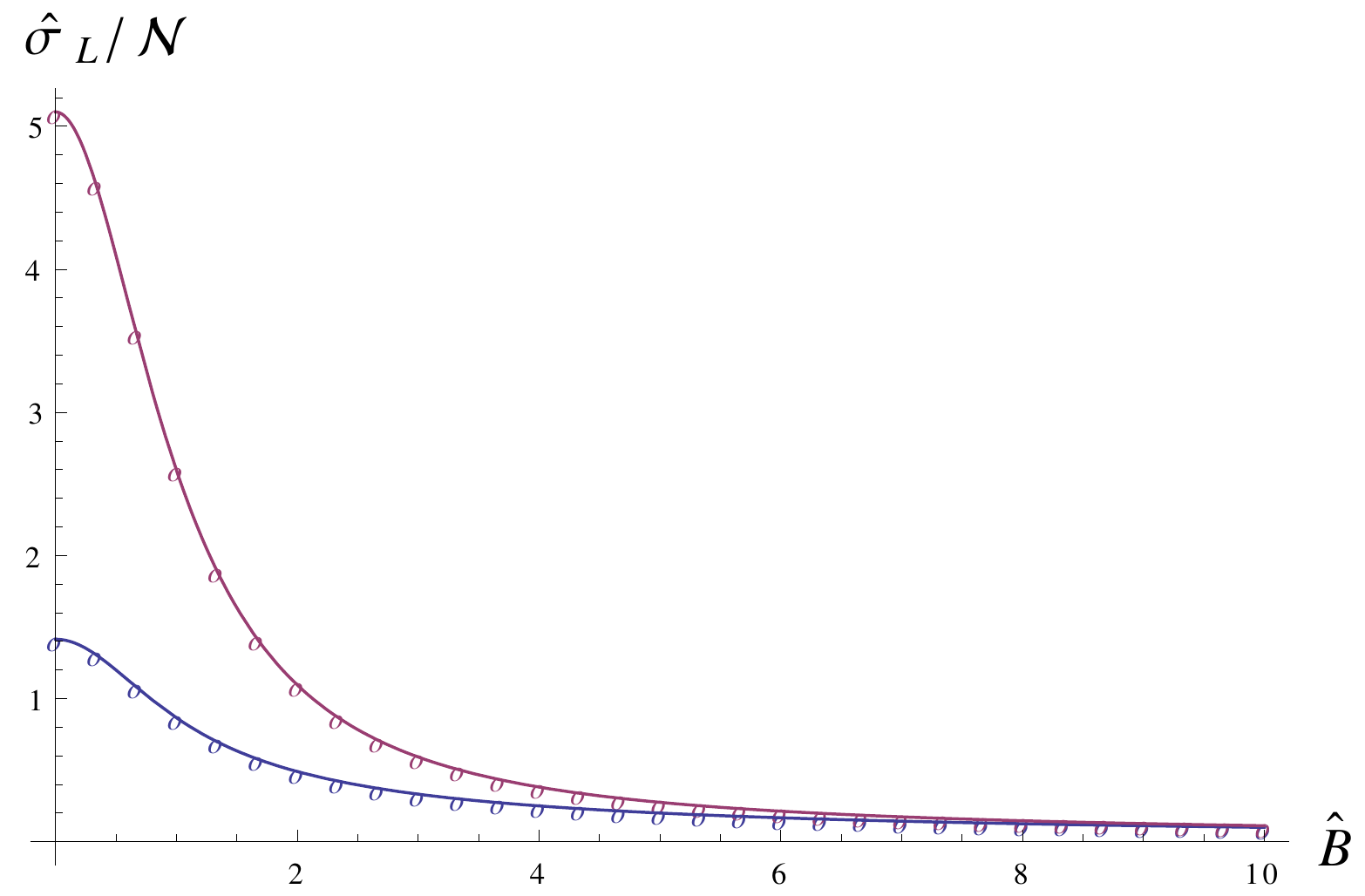}
 \qquad\qquad
  \includegraphics[width=0.40\textwidth]{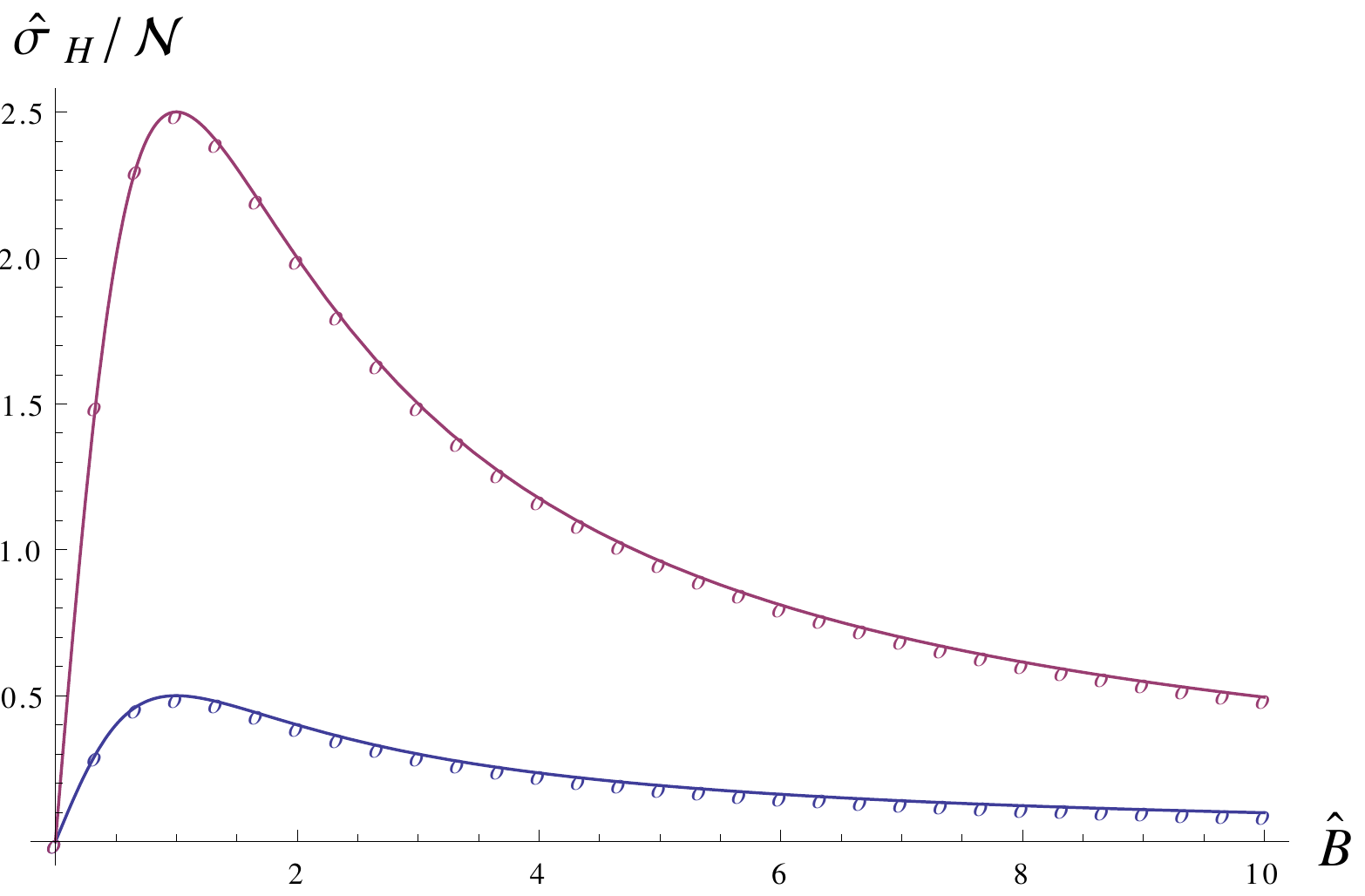}
\caption{A comparison of the numerical (points) and the analytical (continuous curves from (\ref{eq:dc})) results for both the longitudinal (left) and Hall (right)  DC conductivities of the system with $\hat d=1$ (lower) and $\hat d=5$ (upper).
}
\label{conductivities_B}
\end{figure}

Let us collect together the pieces needed for the Einstein relation, in the presence of the magnetic field. First, the diffusion coefficient appears in \eqref{D_result}. Second, for the susceptibility we need first the chemical potential. This can be computed using \eqref{eq:chem} but now with the magnetic field turned on, \emph{i.e.}, we need to use the expression \eqref{eq:At_prime} for $A'_t$. Susceptibility then follows from $\chi = \mathcal{N}\left(\frac{\partial \mu}{\partial d}\right)_{B,T}^{-1}$. 

However, the situation is less straightforward as both the diffusion coefficient and susceptibility contain integrals that we were not able to evaluate. Luckily, it turns out that these integrals are the same upto a constant coefficient and that they cancel when considering the product of the diffusion coefficient and susceptibility. This gives us the longitudinal DC conductivity
\beq\label{eq:dc}
\sigma_L = D\chi = \mathcal{N}r_H^{\xi(q-2)}\frac{\sqrt{1+{\hat d}^2+{\hat B}^2}}{1+{\hat B}^2} \ .
\eeq
We note that we have also included an alternative derivation of this result in Appendix \ref{Karch-O'Bannon}, which we were able to utilize with probe branes.

It is somewhat surprising that the DC conductivity could have such a simple expression. In addition, when considering the dimensionless expression ${\hat \sigma}_L=\sigma_L\,r_H^{\xi(2-q)}$, there is no dependence on the parameters $q$, $\xi$, or $z$. Furthermore, even though our expression for $D$ became unreliable when $g$ in \eqref{eq:g} was smaller than $0.1$ and even divergent when $g<0$, no such difficulties are present in our current situation. Indeed, comparing our results to numerical calculations (see Fig.~\ref{conductivities_B}), the expression we have in (\ref{eq:dc}) agrees virtually perfectly, even if we go to parameter region where $g<0$.


\section{Conclusions}\label{conclusions}

In this paper, we studied holographic matter with both Lifshitz scaling $z$ and hyperscaling violating exponent $\theta$. We allowed these parameters to take any values with only modest assumptions on their range. Moreover, we did not constrain the spatial dimensionality of the (defect) conformal field theory and thus maintained as generic approach as possible. We aimed at drawing general lessons of what are the universal features shared by different holographic models. An important conclusion is the following. While the holographic matter under study has four different parameters: the spatial dimensionalities of the ambient $p$ and defect field theories $q\leq p$, together with $z$ and $\theta$, only three parameters (say $q,\ z$, and $\xi\equiv 1-\theta/p$) were needed for complete description of all physics processes. 

We extracted several key properties of the cold, dense matter as modeled holographically by adding probe D-branes in the background of the most generic metric possessing the parameters $z\ne 1$ and $\theta\ne 0$. However, as such metrics are not yet derived from first principles using concrete brane constructions (except for a few exceptions), our work should be regarded as string inspired. In particular, we did not allow for a non-trivial dilaton in the background geometry. This, however, has the advantage that it is much more straightforward to apply the rules of holographic dictionary for the Lifshitz backgrounds \cite{Taylor:2015glc}.

We worked out the standard thermodynamics, and put special emphasis on the thermodynamic first sound. We then focused on the fluctuations of the probe D-branes and computed the quasi-normal mode spectrum. We carefully contrasted the results that we obtained for the collective modes ({\emph{e.g.}}, zero sound and diffusion mode) to those from the background thermodynamics. Most of the results that we obtained were completely novel, in particular all $B\ne 0$ results are original.

The latter part of the paper focused on holographic matter with fractional spin. That is, we restricted the field theory to reside in 2+1 dimensions and studied a dense system of anyons both at finite magnetic field and temperature. The holographic realization of rendering the standard charge carriers to anyons is to perform alternative quantization for the bulk gauge field. We briefly reviewed this procedure, but quickly turned to analyzing the collective excitations of the anyonic fluid. Our focus was again on the diffusion mode and the zero sound and their behavior under varying $z$ and $\theta$. One of the great successes was to show that the Einstein relation holds, no matter the parameter values. 

There are several avenues where our work could be directed in the future.  One of the most pressing issues is to try to come up with an approximation scheme to analytically capture temperature corrections for the dispersion relation for the zero sound. This has not been achieved in any holographic model. Another important question is to understand the effect of the backreaction of the charge density on the background geometry. For example, it is currently an open question if the zero sound exists in such settings, {\emph{e.g.}}, in electron star/cloud geometries \cite{Hartnoll:2010gu,Hartnoll:2010ik,Puletti:2010de}. We hope to  give a definite answer to this puzzle in near future \cite{NJcloud}.


\vspace{0.5cm}

{\bf \large Acknowledgments}
We thank Daniel Are\'an, Georgios Itsios, and Tobias Zingg for discussions. 
N.~J. and J.~J. are supported in part by the Academy of Finland grant no. 1268023. 
A.~V.~R. is funded by the Spanish grant FPA2011-22594, by the Consolider-Ingenio 2010 Programme CPAN (CSD2007-00042), by Xunta de
Galicia (Conselleria de Educaci\'on, grant INCITE09-206-121-PR and grant PGIDIT10PXIB206075PR), and by FEDER. 
N.~J. wishes to thank Technion for warm hospitality while this work was being finished.

\appendix

\vskip 1cm
\renewcommand{\theequation}{\rm{A}.\arabic{equation}}
\setcounter{equation}{0}

\section{Some useful integrals}\label{appendix:calculations}
Let us collect in this appendix some integrals which are useful in the analysis of the collective excitations of the Lifshitz matter. First of all, we define the integral $I_{\lambda_1,\lambda_2}(r)$ as:
\beq
I_{\lambda_1,\lambda_2}(r)\,\equiv\,\int_{r}^{\infty}\,
{\rho^{\lambda_1}\,d\rho\over (\rho^{\lambda_2}+d^2)^{{1\over 2}}}\,\,.
\label{I_lambda12_def}
\eeq
This integral can be explicitly performed in terms of the hypergeometric function:
\beq
I_{\lambda_1,\lambda_2}(r)\,=\,{2\over \lambda_2-2\lambda_1-2}\,
r^{1+\lambda_1-{\lambda_2\over 2}}\,
F\Big({1\over 2}, {1\over 2}-{\lambda_1+1\over\lambda_2};{3\over 2}-{\lambda_1+1\over\lambda_2}
;-{d^2\over r^{\lambda_2}}\Big) \ .
\label{I_lambda12_value}
\eeq
For small $r$, assuming that $\lambda_2$ and $\lambda_1+1$ are positive, we have the expansion:
\beq
I_{\lambda_1,\lambda_2}(r)={1\over \lambda_2}\,
B\Big({\lambda_1+1\over \lambda_2}, {1\over 2}-{\lambda_1+1\over \lambda_2}\,
\Big)\,d^{2{\lambda_1+1\over \lambda_2}-1}\,-\,{r^{\lambda_1+1}\over (\lambda_1+1)d}+\ldots \ .
\label{I_lambda12_expansion}
\eeq
Let us next define $J_{\lambda_1,\lambda_2}(r)$ in the form:
\beq
J_{\lambda_1,\lambda_2}(r)\,\equiv\,\int_{r}^{\infty}\,
{\rho^{\lambda_1}\,d\rho\over (\rho^{\lambda_2}+d^2)^{{3\over 2}}}\,\,,
\label{J_integral_definition}
\eeq
which can also be computed explicitly:
\beq
J_{\lambda_1,\lambda_2}(r)\,=\,{2\over 3\lambda_2-2\lambda_1-2}\,
r^{1+\lambda_1-{3\lambda_2\over 2}}\,
F\Big({3\over 2}, {3\over 2}-{\lambda_1+1\over\lambda_2};{5\over 2}-{\lambda_1+1\over\lambda_2}
;-{d^2\over r^{\lambda_2}}\Big)\,\,.
\label{J_value}
\eeq
For small $r$, when  $\lambda_2$ and $\lambda_1+1$ are both positive, we can   expand 
$J_{\lambda_1,\lambda_2}(r)$ as:
\beq
J_{\lambda_1,\lambda_2}(r)={1\over \lambda_2}\,
B\Big({\lambda_1+1\over \lambda_2}, {3\over 2}-{\lambda_1+1\over \lambda_2}\,
\Big)\,d^{2{\lambda_1+1\over \lambda_2}-3}\,-\,{r^{\lambda_1+1}\over (\lambda_1+1)d^3}+\ldots \ .
\eeq

\vskip 1cm
\renewcommand{\theequation}{\rm{B}.\arabic{equation}}
\setcounter{equation}{0}

\section{The Wronskian method}
\label{wronskian}

Let us now solve the inhomogeneous equation (\ref{inhomo_y_eq}) by applying the Wronskian method.  First, we define a new function ${\cal Y}$ as:
\beq
{\cal Y}\,\equiv r^{{z\over 2}-\xi}\,\,y(r)\,\,,
\eeq
and a new independent variable $x$ as:
\beq
x\,\equiv {\omega\over z}\,r^{-z}\,\,.
\eeq
In terms of ${\cal Y}(x)$, the inhomogeneous equation (\ref{inhomo_y_eq}) becomes:
\beq
{d^2\,{\cal Y}\over d\,x^2}\,+\,{1\over x}\,{d\,{\cal Y}\over dx}\,+\,
\Big(1\,-\,{\nu^2\over x^2}\Big)\,{\cal Y}\,=\,d_z\,x^{-2\nu}\,H_{\nu}^{(1)}(x)\,\,,
\label{eq_cal_Y}
\eeq
where $\nu$ and $d_z$ are the following constants:
\beq
\nu={1\over 2}-{\xi\over z}\,\,,
\qquad\qquad
d_z\,=\,-{4\xi \over \omega}\,\Big({z\over \omega}\Big)^{-2\nu}\,\,.
\label{nu_dz}
\eeq
It is actually more convenient to write (\ref{eq_cal_Y}) in terms of Hankel functions of index 
\beq
\bar\nu=-\nu\,=\,{\xi\over z}-{1\over 2}\,\,.
\eeq
Taking into account that
\beq
H_{-\nu}^{(1)}(x)\,=\,e^{i\pi \nu}\,H_{\nu}^{(1)}(x)\,\,,
\qquad\qquad
H_{-\nu}^{(2)}(x)\,=\,e^{-i\pi \nu}\,H_{\nu}^{(2)}(x)\,\,.
\eeq
We can rewrite (\ref{eq_cal_Y})  as:
\beq
{d^2\,{\cal Y}\over d\,x^2}\,+\,{1\over x}\,{d\,{\cal Y}\over dx}\,+\,
\Big(1\,-\,{\bar \nu^2\over x^2}\Big)\,{\cal Y}\,=\,f(x)
\label{eq_cal_Y_bar_nu}
\eeq
where $f(x)$ is the function:
\beq
f(x)\,=\,\bar d_z\,x^{2\bar\nu}\,H_{\bar\nu}^{(1)}(x)\,\,,
\eeq
with
\beq
\bar d_z\,=\,e^{i\pi\bar \nu}\,d_z\,\,.
\eeq
 Let ${\cal Y}_1(x)$ and ${\cal Y}_2(x)$ be two independent solutions of (\ref{eq_cal_Y}) with $f(x)=0$. Then, the solution of 
(\ref{eq_cal_Y}) for $f(x)\not=0$ can be written as:
\beq
{\cal Y}(x)\,=\,I_1(x)\,{\cal Y}_1(x)\,+\,I_2(x)\,{\cal Y}_2(x)\,\,,
\label{particular_sol}
\eeq
where $I_1(x)$ and $I_2(x)$ are the following indefinite integrals:
\beq
I_1(x)\,=\,-\int {{\cal Y}_2(x)\,f(x)\over W({\cal Y}_1, {\cal Y}_2)}\,dx\,\,,
\qquad\qquad
I_2(x)\,=\,\int {{\cal Y}_1(x)\,f(x)\over W({\cal Y}_1, {\cal Y}_2)}\,dx\,\,,
\eeq
and $ W({\cal Y}_1, {\cal Y}_2)$ is the Wronskian function of ${\cal Y}_1$ and ${\cal Y}_2$:
\beq
W({\cal Y}_1, {\cal Y}_2)\,=\,{\cal Y}_1\, {\cal Y}_2{'}\,-\,{\cal Y}_2\, {\cal Y}_1{'}\,\,.
\eeq
The homogeneous version of (\ref{eq_cal_Y}) is just the Bessel equation. Therefore, we can take the Hankel functions of index $\nu_p$ as the two independent solutions 
${\cal Y}_1$ and  ${\cal Y}_2$:
\beq
{\cal Y}_i(x)\,=\,H_{\bar\nu}^{(i)}(x)\,\,,
\qquad\qquad
(i=1,2)\,\,.
\eeq
The Wronskian of two Hankel functions is rather simple, namely:
\beq
W(H_{\bar\nu}^{(1)}(x), H_{\bar\nu}^{(2)}(x))\,=\,-{4i\over \pi x}\,\,.
\eeq
Therefore, $I_1(x)$ and $I_2(x)$ are given by:
\bear
&&I_1(x)\,=\,-i{\pi \bar d_z\over 4}\,\int x^{2\bar\nu+1}\,
H_{\bar\nu}^{(2)}(x)\,H_{\bar\nu}^{(1)}(x)\,dx\rc
&&I_2(x)\,=\,i{\pi \bar d_z\over 4}\,\int x^{2\bar\nu+1}\,
H_{\bar\nu}^{(1)}(x)\,H_{\bar\nu}^{(1)}(x)\,dx\,\,. 
\eear
Taking into account that (for $\nu+\mu\not=1$):
\beq
 \int x^{\mu+\nu+1}\,
H_{\mu}^{(\alpha)}(x)\,H_{\nu}^{(\beta)}(x)\,dx\,=\,
{x^{\mu+\nu+2}\over 2(\mu+\nu+1)}\,\Big[
H_{\mu}^{(\alpha)}(x)\,H_{\nu}^{(\beta)}(x)\,+\,
H_{\mu+1}^{(\alpha)}(x)\,H_{\nu+1}^{(\beta)}(x)\,\Big]\,\,,
\eeq
we get that $I_1(x)$ and $I_2(x)$ are given by:
\bear
I_1(x) & =&-i{\pi \bar d_z\over 8}\,
{x^{2\bar\nu+2}\over 2\bar\nu+1}
\Big[\,H_{\bar\nu}^{(2)}(x)\,H_{\bar\nu}^{(1)}(x)\,+\,
H_{\bar\nu+1}^{(2)}(x)\,H_{\bar\nu}^{(1)}(x)\,\Big] \rc
I_2(x)&=&\,i{\pi \bar d_z\over 8}\,
{x^{2\bar\nu+2}\over 2\bar\nu+1}
\Big[\,H_{\bar\nu}^{(1)}(x)\,H_{\bar\nu}^{(1)}(x)\,+\,
H_{\bar\nu+1}^{(1)}(x)\,H_{\bar\nu+1}^{(1)}(x)\,\Big]\,\,. 
\eear
Let us plug these values in (\ref{particular_sol}) and use the following property of the Hankel functions:
\beq
H_{\bar\nu+1}^{(1)}(x)\, H_{\bar\nu}^{(2)}(x)\,-\,
H_{\bar\nu}^{(1)}(x)\, H_{\bar\nu+1}^{(2)}(x)\,=\,-{4i\over x}\,\,.
\eeq
We get  that ${\cal Y}$ is given by:
\beq
{\cal Y}\,=\,{\bar d_z\over 2(2\bar\nu+1)}\,x^{2\bar\nu+1}\,
H_{\bar\nu+1}^{(1)}(x)\,=\,r^{-2\xi}\,
H_{-{\xi\over z}-{1\over 2}}^{(1)}\Big({\omega\over z r^z}\Big)\,\,.
\eeq
Therefore, we get the following solution for $y(r)$:
\beq
y(r)\,=\,r^{-\xi-{z\over 2}}\,
H_{-{\xi\over z}-{1\over 2}}^{(1)}\Big({\omega\over z r^z}\Big)\,\,.
\eeq

\vskip 1cm
\renewcommand{\theequation}{\rm{C}.\arabic{equation}}
\setcounter{equation}{0}

\section{Transverse correlators at finite temperature}\label{appendix:transverse}

To fully compute the effect of mixed boundary conditions, we need to solve the equations of motion for the transverse field $a_y$ at finite temperature. It turns out, however, that the effect of the magnetic field is rather large and the kind of approximation scheme that we pursued did not provide satisfactorily accurate results. Therefore, we will set $B=0$ in this appendix. 

First, we expand equation \eqref{eq:ayeom} near the horizon. The coefficient of the term multiplying $a_y$ is same as the corresponding term in \eqref{eq:Eeom}. For the derivative term, we have
\beq
\partial_r\log\left[\frac{\sqrt{|g_{tt}|}}{\sqrt{g_{rr}}}f_p\frac{\sqrt{H+d^2}}{g_{xx}}\right] = \frac{1}{r-r_H}+d_1+\mathcal{O}(r-r_H) \ ,
\eeq
where
\beq
d_1 = \frac{\xi  \left(\frac{2 q r_H^{2 \xi  q}}{d^2+r_H^{2 \xi  q}}-p-4\right)+z+1}{2 r_H}\ .
\eeq
Once again, we solve this equation with a Frobenius series of the form $a_y = (r-r_H)^{\alpha}(1+\beta (r-r_H)+\ldots)$. We see that the $\alpha$ here is equal to the one in \eqref{eq:alpha}. Coefficient $\beta$ is solved as before and, in the hydrodynamic regime with $k^2\sim\omega\sim \epsilon^2$, we get
\beq
\beta\approx \frac{k^2 r_H^{2 \xi  q-3}}{(\xi  p+z) \left(d^2+r_H^{2 \xi  q}\right)}+i \omega \frac{ r_H^{-z-1} \left((d^2+r_H^{2 \xi  q}) (-\xi  (p+4)+z+1)+2\xi qr_H^{2 \xi  q} \right)}{2 (\xi  p+z) \left(d^2+r_H^{2 \xi  q}\right)} \ .
\eeq

Let us take the opposite order. We write the equation (\ref{eq:ayeom}) in the form
\beq
a_y''+\frac{G'}{G}a_y'+k^2Qa_y=0 \ ,
\eeq
where the coefficients in the low frequency limit are
\bear
G &=& r^{-2\xi+z+1} (r^{2\xi q}+d^2)^{{1\over 2}} f_p \\
Q &=& \frac{-r^{2q\xi-4}}{(d^2+r^{2q\xi})f_p} \ .
\eear
As we wish to match our solution to our previous expansion, we introduce the function $\alpha_y(r)$,
\beq
a_y = F(r)\alpha_y(r) \ ,
\eeq 
where $F(r) = (r-r_H)^{\alpha}$. The $\alpha$ in the expression is the same as before, {\emph{i.e.}}, $\alpha\sim \epsilon^2$. The function $\alpha_y$ should now be finite near the horizon. The equation for $\alpha_y$ becomes
\beq
\alpha_y''+\left(\frac{G'}{G}+2\frac{F'}{F}\right)\alpha_y'+(k^2 Q+\omega P)\alpha_y=0\ ,
\eeq
where 
\beq
\omega P = \frac{F''}{F}+\frac{G'}{G}\frac{F'}{F} \ .
\eeq
Due to $\alpha$, we note that $F'\sim F''\sim \epsilon^2$. 

We now expand $\alpha_y$ as a series in $\epsilon$, $\alpha_y(r) = \alpha_0(r)+\alpha_1(r)+\ldots$, where $\alpha_n\propto \epsilon^{2n}$. The equation for $\alpha_0$ becomes
\beq
\alpha_0''+\frac{G'}{G}\alpha_0'=0
\eeq
which is solved by
\beq
\alpha_0' = \frac{c_0}{G(r)}\ .
\eeq
This solution diverges when approaching the horizon unless $c_0=0$. Thus, we must have $\alpha_0=\rm{constant}$.

The equation for $\alpha_1$ is
\beq
\alpha_1''+\frac{G'}{G}\alpha_1'=\alpha_0(-k^2 Q-\omega P) \ .
\eeq
We see that the homogeneous part is the same as before, thus we use the method of variation of constants with
\beq
\alpha_1' = \alpha_0\frac{\Lambda(r)}{G(r)} \ ,
\eeq
which leads to
\beq
\Lambda'(r) = G\left[-k^2Q-\omega P\right]=-k^2 G\,Q-\omega\partial_r(G\partial_r\log F) \ ,
\eeq
where the last step is valid upto order $\epsilon^2$. This is now easily integrated, 
\bear
\Lambda &=& -G(r)\frac{F'}{F}-k^2\int\limits_{r_H}^{r}Q(\rho)G(\rho){\rm d}\rho-c_1 \\
&=& -\alpha \frac{G}{r-r_H}+k^2\int\limits_{r_H}^{r}\frac{\rho^{2\xi(q-1)+z-3}}{\sqrt{d^2+\rho^{2\xi q}}}{\rm d}\rho-c_1 \ ,
\eear
where $c_1$ is a constant which we will determine later on. For future use in this section, we will denote the integral in the last step as $\mathcal{I}(r)$. Thus, we find that $\alpha_1'$ is
\beq
\alpha_1' = -\alpha_0\left(\frac{\alpha}{r-r_H}+\frac{c_1}{G(r)}\right)+k^2\frac{\mathcal{I}(r)}{G(r)} \ .
\eeq

We require that $\alpha_1$ be finite at the horizon. We do not need to consider the $\mathcal{I}(r)$ term as the integral vanishes linearly, taking care of the divergence caused by $1/G(r)$. The expansion of $1/G$ near the horizon is
\beq
\frac{1}{G}=\frac{r_H^{2\xi-z}}{\left(r-r_H\right)(\xi p+z)\sqrt{d^2+r_H^{2\xi q}}}-\frac{r_H^{2\xi-z-1}\left((-\xi(p+4)+z+1)\left(d^2+r_H^{2\xi q}\right)+2\xi q r_H^{2\xi q}\right)}{2(\xi p+z)\left(d^2+r_H^{2\xi q}\right)^{3/2}}+\ldots \ ,
\eeq
from which we can see that
\beq
c_1 = -\alpha  (\xi  p+z) r_H^{z-2 \xi } \sqrt{d^2+r_H^{2 \xi  q}}=i \omega  r_H^{-2 \xi } \sqrt{d^2+r_H^{2 \xi  q}} \ ,
\eeq
for $\alpha_1$ to be finite at the horizon.
Thus,
\beq
\alpha_y' = -\alpha_0\alpha\left(\frac{1}{r-r_H}-\frac{(\xi  p+z) r_H^{z-2 \xi } \sqrt{d^2+r_H^{2 \xi  q}} }{G}\right)+\alpha_0k^2\mathcal{I}(r) \ .
\eeq

We should now check that our two solutions match. As the IR expansion of $\mathcal{I}(r)$ is,
\beq
\mathcal{I}(r)=\frac{r_H^{2\xi(q-1)+z-3}}{\sqrt{d^2+r_H^{2\xi q}}}(r-r_H)+\ldots \ ,
\eeq
we can calculate that
\beq
\alpha_y'(r=r_H) = \alpha_0\beta \ ,
\eeq
as it should be.
Furthermore, 
\bear
a_y &=& F(r)(\alpha_0+\alpha_1)+\mathcal{O}(\epsilon^4)=\alpha_0+\alpha_1+\alpha_0\alpha\log(r-r_H)+\mathcal{O}(\epsilon^4)\\
a_y'&=& \frac{\alpha_0}{G(r)}\left(k^2\mathcal{I}(r)-i\omega r_H^{-2\xi}\sqrt{d^2+r_H^{2\xi q}}\right)+\mathcal{O}(\epsilon^4) \ .
\eear

For the UV limit, we need to evaluate $\mathcal{I}(r)$ as $r\to \infty$. The integral can be evaluated analytically by using the formula \eqref{I_lambda12_def} in Appendix \ref{appendix:calculations}:
\beq
\mathcal{I}(r\to \infty) =I_{\lambda_1,\lambda_2}(r_H)= \frac{r_H^{\xi  (q-2)+z-2} \, _2F_1\left(\frac{1}{2},-\frac{z+(q-2) \xi -2}{2 q \xi };\frac{-z+(q+2) \xi +2}{2 q \xi };-d^2 r_H^{-2 q \xi }\right)}{2-\xi  (q-2)-z} \ .
\eeq
Thus, the UV limit of $a_y'$ is
\bear
a_y' &=& \alpha_0 r^{-\xi  (q-2)-z-1}\left(k^2\mathcal{I}(\infty)-i\omega r_H^{-2\xi}\sqrt{d^2+r_H^{2\xi q}}   \right) \label{eq:defofZ}\\
&=& \alpha_0 r^{-\xi  (q-2)-z-1} Z(\omega,k) \ ,
\eear
where we have defined function $Z(\omega,k)$ to shorten the notation for later use.

Finally, we compute the two-point function $\langle J_y(-k)J_y(k)\rangle$. In the action, $a_y$ appears only in the form
\beq
\int {\rm d}^{q+1}k{\rm d}r\mathcal{F}(a_y')^2 \ ,
\eeq
where
\beq
\mathcal{F} = -\frac{\mathcal{N}}{2}\frac{\sqrt{g_{rr}|g_{tt}|}}{\sqrt{H+d^2}}H\mathcal{G}^{yy}\mathcal{G}^{rr}=-\frac{\mathcal{N}}{2}G(r) \ .
\eeq
Performing a partial integration along the radial coordinate and using the equations of motion, we can transform the integral to a boundary one
\beq
S_{\rm{on-shell}}(a_y) = \int{\rm d}^{q+1}k\mathcal{F}a_y a_y'|_{r\to \infty}\ .
\eeq

As per the usual prescription, varying twice with respect to the boundary value $\alpha_0(k)$, we get the two-point function of the current $J_y$, {\emph{i.e.}},
\beq
\langle J_y(-k)J_y(k)\rangle = -\mathcal{N}\left(k^2 \mathcal{I}(\infty)-i\omega r_H^{-2\xi}\sqrt{d^2+r_H^{2\xi q}}\right)\ .
\eeq
We notice that the two-point function has a zero at
\bear
\omega = -i D^*k^2, \quad D^* = \frac{r_H^{z-2} \, _2F_1\left(\frac{1}{2},-\frac{z+(q-2) \xi -2}{2 q \xi };\frac{-z+(q+2) \xi +2}{2 q \xi };-d^2 r_H^{-2 q \xi
   }\right)}{(2-\xi  (q-2)-z)\sqrt{1+d^2r_H^{-2\xi q}}}\ .\label{dstar}
\eear
We emphasize that this is not a diffusion mode. The use of symbol $D^*$ will be useful when considering alternative quantization.

\vskip 1cm
\renewcommand{\theequation}{\rm{D}.\arabic{equation}}
\setcounter{equation}{0}

\section{DC conductivity from Karch-O'Bannon}
\label{Karch-O'Bannon}

It is not surprising that there are other methods of computing DC conductivity which do not require explicit computations of dispersion relations and two-point functions. In particular, for probe branes, there is a powerful non-linear method developed by Karch-O'Bannon \cite{Karch:2007pd}. We will apply this method to solve both the longitudinal and Hall conductivity of the system in the presence of a magnetic field.

We will consider background electric field and the corresponding currents by turning on additional gauge fields in our original DBI action. We need an electric field in the $x$ direction, denoted by $e$, and the corresponding currents, $j_x$ and $j_y$, will be encoded in the radial components of the gauge field. Thus, the additional non-zero terms are
\beq
F_{tx} = e\quad , \quad F_{rx} = a_x'(r)\quad , \quad F_{ry} = a_{y}'(r) \ .
\eeq

The DBI action in this case takes the form
\be
 S_{DBI}=-\mathcal{N} \int {\rm d}^{q+1}x{\rm d}r\,r^{-3+z+(2+q)\xi}\sqrt Y \ ,
\ee
where
\bea
Y & = &  \left(1+\frac{B^2}{r^{4\xi}}-\frac{e^2r^{2-2z-4\xi}}{f_p}\right)-r^{6-2z-4\xi}\left(1+\frac{B^2}{r^{4\xi}}\right){A_t'}^2\\
&& + f_p r^{4-4\xi}\left(1-\frac{e^2r^{2-2z-4\xi}}{f_p}\right)a'^2_y + f_pr^{4-4\xi}a'^2_x - 2Be r^{6-2z-8\xi}A_t' a_y' \ .
\eear
When extremizing the action, we see that the gauge fields are all cyclic variables, which allows us to introduce three constants of motion, $d$, $j_x$, and $j_y$ corresponding to $A_t$, $a_x$, and $a_y$, respectively. Using this to our advantage, we can write down the equations of motion
\bear
r^{3-z+(q-6)\xi}\left(B e a_y'+(B^2+r^{4\xi})A_t'\right) &=& d\sqrt Y  \\
a_x'f_p r^{1+z+(q-2)\xi} &=& -j_x \sqrt Y  \\
r^{1-z+(q-6)\xi}\left(   (r^{2(z+2\xi)}f_p-e^2r^2) a_y' - eB r^2 A_t' \right) &=& -j_y \sqrt Y \ .
\eear

We reshuffle the above equations to obtain expressions for the gauge fields
\bear
a_x' &=& \frac{-j_x\sqrt{Y}}{f_p r^{1+z+(q-2)\xi}}\\
a_y' &=& \frac{\sqrt{Y}(B d e -(B^2+r^{4\xi})j_y)}{f_pr^{1+z+(q+2)\xi}\left(1+\frac{B^2}{r^{4\xi}}-\frac{e^2 r^{2-2z-4\xi}}{f_p}\right)}\\
A_t' &=& \frac{\sqrt{Y}\left( f_p d r^{2(z+2\xi)} (1-\frac{e^2}{f_p}r^{2(1+z+2\xi)})+ B e r^2 j_y\right)}{f_pr^{3+z+(q+2)\xi}\left(1+\frac{B^2}{r^{4\xi}}-\frac{e^2 r^{2-2z-4\xi}}{f_p}\right)} \ .
\eear
and then substitute these into our expression of $Y$ and then solve $Y$ in terms of $j_x$, $j_y$, and $d$:
\bear
\sqrt Y &=&\frac{r^{(q-2)\xi+4\xi+2z}f_p(1+\frac{B^2}{r^{4\xi}}-\frac{e^2 r^{2-2z-4\xi}}{f_p})}{\sqrt{X}} \\
X &=& r^{2(z+2\xi)}f_p\left(1+\frac{B^2}{r^{4\xi}}-\frac{e^2 r^{2-2z-4\xi}}{f_p} \right)(r^{2z}(d^2+r^{2\xi q})f_p-(j_x^2+j_y^2)r^2) \\
&&\quad -(ej_y r^2-B d r^{2z}f_p)^2\ .
\eear

Now we take a closer look at $X$. We see that it cannot go to zero or have negative values at any point as this would cause a divergence of the gauge fields. We see that the first brackets in the first term has a zero near the horizon as long as $e$ is non-zero. This zero corresponds to the location of the pseudo-horizon. We label this radius with $r_*$. We also require that the term in the second brackets of the first term and the last term vanishes at $r_*$. These give us two new conditions which we can use to determine $j_x$ and $j_y$. To get analytic results, we expand all results to first non-trivial order in $e$. The conditions are
\bear
e^2 r^2 &=& r^{2 z} (B^2 + r^{4 \xi}) f_p(r_*)\\
(j_x^2 + j_y^2) r_*^2 &=& r_*^{2 z} (d^2 + r_*^{2 q \xi}) f_p(r_*) \\
e jy r_*^2 &=& B d r_*^{2 z} f_p(r_*) \ .
\eear

Solving for $r_*$ in first order of $e$ and evaluating $f_p$ at $r_*$, we get
\bear
r_* &=& r_H + \frac{e^2 r_H^{3-2z}}{(B^2+r_H^{4\xi}(z+p\xi)}\\
f_p(r_*) &=& \frac{e^2 r_H^{2-2z}}{B^2+r_H^{4\xi}} \ .
\eear
Solving for $j_y$ and $j_x$ at first order in $e$, we get
\bear
j_y &=& \frac{B d e}{B^2 + r_H^{4\xi}} \\
j_x &=& e\frac{d^2 r_H^{4\xi}+r_H^{2q\xi}(B^2+r_H^{4\xi})}{B^2+r_H^{4\xi}} \ .
\eear

Similarly to the relation between $d$ and the physical charge density $\rho$, we must also multiply $j_x$ and $j_y$ by ${\cal N}$ to obtain the conductivity of this system. Thus, we get for the longitudinal and Hall conductivity
\beq
\sigma_L = \mathcal{N}r_H^{(q-2)\xi}\frac{\sqrt{1+{\hat B}^2+{\hat d}^2}}{1+\hat{B}^2}\quad , \quad \sigma_H= \mathcal{N} r_H^{(q-2)\xi}\frac{{\hat B} {\hat d}}{1+ {\hat B}^2}\ ,
\eeq
where we have used the scaled variables in accordance with our previous results for DC conductivities. We find that our result for longitudinal conductivity exactly matches with our result obtained from the Einstein relation.


\end{document}